\documentclass[twocolumn,showpacs,preprintnumbers,amsmath,amssymb]{revtex4}
\usepackage{epsfig}
\usepackage{graphicx}
\usepackage{dcolumn}
\usepackage{bm}
\usepackage{pstricks}

\newcommand{\remove}[1]{}
\newcommand{\dd}{\mathrm{d}}
\newcommand{\rx}{{\rm x}}

\def\be{\begin{equation}}
\def\ee{\end{equation}}

\newcommand{\beq}{\begin{equation}}
\newcommand{\eeq}{\end{equation}}
\newcommand{\beqa}{\begin{eqnarray}}
\newcommand{\eeqa}{\end{eqnarray}}

\newcommand{\bea}{\begin{array}}
\newcommand{\ea}{\end{array}}

\begin{document}

\title{Cosmological tests of modified gravity: \\ constraints on  $F(R)$ theories from the galaxy clustering ratio}

\author{Julien Bel$^1$, Philippe Brax$^2$, Christian Marinoni$^3$ \& Patrick Valageas$^2$}
\affiliation{$^1$ INAF - Osservatorio Astronomico di Brera, Via Brera 28, 20122 Milano, via E. Bianchi 46, 23807 Merate, Italy\\
$^2$ Institut de Physique Th\'eorique, CEA, IPhT, F-91191 Gif-sur-Yvette, C\'edex, France\\
CNRS, URA 2306, F-91191 Gif-sur-Yvette, C\'edex, France\\
$^3$ Aix Marseille Universit\'e, Universit\'e de Toulon, CNRS, CPT UMR 7332, 13288, Marseille, France\\
Institut Universitaire de France, 103, bd. Saint-Michel, F-75005 Paris, France}

\date{\today}

\begin{abstract}
The clustering ratio $\eta$, a large-scale structure observable originally designed to constrain the shape of the power spectrum of matter density fluctuations,  is shown to provide  a sensitive  probe of  the nature of gravity  in the cosmological regime.  We apply this analysis to $F(R)$ theories of gravity using the luminous red galaxy (LRG) sample extracted from the spectroscopic Sloan Digital Sky Survey (SDSS) data release 7 and 10 catalogues.
We find that General Relativity (GR), complemented with a Friedmann-Robertson-Walker (FRW) cosmological model with parameters fixed by the Planck satellite, describes extremely well the clustering of galaxies up to $z\sim 0.6$. On large cosmic scales, the absolute amplitude of deviations from GR, $|f_{R_0 }|$, is constrained to be smaller than $4.6 \times 10^{-5}$ at the $95\%$ confidence level.    This bound makes cosmological probes of gravity almost competitive with  the sensitivity of Solar System tests, although still one order of magnitude less effective than astrophysical tests.
We also extrapolate our results to future large surveys like Euclid and show that  the astrophysical bound will certainly remain out of reach for such a class of  modified-gravity models that only differ from $\Lambda$CDM at low redshifts.\end{abstract}

\maketitle

Perplexing  observations, such as the accelerated expansion of the universe  (the dark energy phenomenon  \cite{SCP,SST}),
the rotation curves of galaxies or the gravitational  lensing from  large clusters of galaxies (the dark matter phenomenon \cite{os,fw}),
seem to point towards new phenomena beyond the physics already tested in the laboratory or in the Solar System. In particular, they may be associated with departures
from Einstein's theory of gravity \cite{revMG,PSM}.

Powerful tests on  Solar System \cite{Williams:2012nc} and astrophysical \cite{Jain:2012tn,Vikram:2013uba} scales, however,
seem to confirm  GR predictions and  impose stringent limits on  modified gravity theories.
As a consequence, realistic alternative  theories should incorporate  non-linear
mechanisms that ensure convergence to General Relativity on small scales and high-density environments.
In particular, the fifth force which emerges as a generic prediction of modified gravity models,  obtained  by adding a single
scalar degree of freedom $\varphi$ to Einstein's  equations, can be effectively  {\it screened} either by
the  Vainshtein  \cite{Vain},  the Damour-Polyakov \cite{DP}, or  the chameleon \cite{KW,KW2,BBDKW} mechanisms.
Models with the chameleon property converge to GR on cosmological scales too.
As a consequence, intermediate, mildly non-linear scales ($\sim 10 h^{-1}$Mpc)  appear as a unique window of opportunity for detecting
possible deviations from GR in a  large class of models.

Modified gravity models must, in a first approximation,
reproduce the smooth  background expansion history of the standard model of cosmology, the   $\Lambda$-Cold Dark Matter ($\Lambda$CDM) paradigm.
To distinguish  and falsify various competing  gravitational proposals it is thus necessary
to analyze  characteristic observables  of  the perturbed sector of the model.
Indeed,  to lowest order in cosmological perturbations, non-standard gravitational
scenarios effectively result in a time- and scale-dependent modification of
Newton's constant, that is,  in a distortion of the dynamical and the statistical properties
of characteristic clustering quantities  such as the  power spectrum  \cite{BBPST, BV2, HRFS} and the growing mode
$D_+(k,z)$ [and its logarithmic derivative, the growth rate  ${\rm f}(k,z)=\partial \ln D_+/\partial\ln a$] of  linear matter perturbations \cite{LSS, PSM}.

Gravity tests on cosmological scales are still far from reaching  the precision  achieved with  small-scales experiments.
Several observational shortcomings  affect  standard probes such as, for example,
cosmic shear in weak lensing maps \cite{wl,Tsujikawa2008}, redshift-space distortions \cite{sam,Guzzo2008}, and galaxy clustering \cite{BBPST,BV2,HRFS,Pogosian2008,Oyaizu2008}.
Their main observable, the growth rate of dark matter ${\rm f}$, for example,  cannot be estimated from data without picking a
particular model, or at least a parameterization, for gravity
\cite{Motta2013, sbm}.  Moreover, although the most generic extensions of GR
predict a scale dependent growth rate ${\rm f}$, devising a method able to measure ${\rm f}(k,z)$ at different scales is  a formidable
observational task \cite{CCD, TKHO}. Additionally, in all the analyses it is assumed that
the bias is the same for both modified and standard gravity models.
This is  a non-trivial ansatz, since not only the growth of structures is expected to be
different in modified theories of gravity, but  the gravitational potential, which plays a key role in the formation of galaxies, and hence in determining their biasing properties, changes.
Finally, most of these probes rely on a precise and challenging  measurement of the mean galaxy density on large cosmic scales \cite{adi}.

Here we  show how  to address  most of these issues  via a  new gravitational probe, the clustering
ratio $\eta$  \cite{bm13a} (hereafter BM14),\cite{bm13b}.  Due to its peculiar definition, this cosmological observable
naturally accounts for possible scale-dependent growth rates of matter fluctuations.  In particular,  on linear scales the $\eta$
amplitude is constant as a function of time in  general smooth dark energy  ($w$-CDM) models but  acquires a characteristic time dependence
for modified gravity models.  Observational information are inferred from the analysis  of  the luminous red galaxy (LRG) sample extracted from the SDSS
data release 7 \cite{sdss7}  as well as from the data release 10 \cite{sdss10}.

In section~\ref{sec:clustering-ratio}, we define the clustering ratio and relate its
measurement to the real-space matter density power spectrum in the quasi-linear regime.
In section~\ref{sec:f(R)-models}, we introduce the models of modified gravity that we
shall be using in this paper, i.e. $F(R)$ theories in the large curvature regime.
These models serve as a first illustration of our results using the clustering ratio.
The same methods can be applied to more complex models, a study which is left for future
work.
We also extract the clustering ratio and its redshift dependence
from the SDSS catalogue.
In section~\ref{sec:Systematics}, we consider the different systematic effects which
hamper the accuracy of the clustering ratio comparison with data. 
This section can be skipped by the reader who is only interested in the applications to modified gravity. 
Finally in section~\ref{sec:constraints}, we obtain constraints on $F(R)$ models when
either keeping the matter fraction $\Omega_{m0}$ fixed or relaxing it using the Planck
prior.
We conclude in section~\ref{sec:Conclusion}.

\section{The galaxy clustering ratio as a gravitational probe}
\label{sec:clustering-ratio}

\subsection{Several clustering ratios and their relations}
\label{sec:Several-clustering-ratios}

The galaxy clustering ratio in redshift space,
\beqa
\eta^{\rm s}_{g}(r,x;z) & \equiv &
\frac{{\xi}^{\rm s}_{{g},x}(r)}{({\sigma}^{s}_{{g},x})^2} 
\label{etag-s-def1} \\
&& \hspace{-2cm} = \frac{\int_0^{\infty} \dd k \, k^2 \int_{-1}^{1}
\dd\mu \, P^{\rm s}_{g}(k,\mu;z) W(k x)^2 \; \frac{\sin(k r)}{k r}}
{\int_0^{\infty} \dd k \, k^2 \int_{-1}^{1} \dd\mu \, P^{\rm s}_{g}(k,\mu;z) W(k x)^2}  ,
\label{etag-s-def}
\eeqa
is  defined as the ratio of the correlation function to the variance of the
redshift-space galaxy over-density field $\delta^{\rm s}_{{g},x}$,
smoothed on a scale $x$ via the filter $W$, where
$W(y)=3 [\sin(y)-y \cos(y)]/y^3$ is the Fourier transform of the unit top-hat,
the specific  filtering scheme adopted by BM14 to smooth data.
Note that the index $g$ labels quantities that are evaluated using
galaxies as opposed to matter, and that the supercript {\it s} indicates when
physical quantities are evaluated in redshift space as opposed to real space.
Furthermore, to simplify the analysis, we consider
\be
\eta^{\rm s}_{g}(n,x;z) \equiv \eta^{\rm s}_{g}(r=n x,x;z)
\label{eta-s-n-def}
\eeq
as a function of the smoothing scale $x$ for a fixed ratio $n$ between the
correlation ($r$) and the smoothing ($x$) scales.

At first  glance, this second order statistic only provides information about the
monopole $P^{\rm s}_{g (0)}$ of the redshift-space power spectrum of the
galaxies.  A key result of BM14, however, was  to show that on quasilinear scales this
second order statistics provides information on a more fundamental and simpler
physical quantity, that is, the power spectrum of matter fluctuations in real space
$P_{g}(k,z)$. Indeed
\be
\eta^{\rm s}_{g}(n,x;z) \approx \eta_{g}(n,x;z) \approx \eta(n,x;z) ,
\label{eta-chain}
\ee
where $\eta(n,x;z)$ is the mass clustering ratio in real space,
\be
\eta(n,x;z) \equiv
\frac{\int_0^{\infty} \dd k \, k^2  \, P(k,z) W(k x)^2 \; \frac{\sin(k n x)}{k n x}}
{\int_0^{\infty} \dd k \, k^2 \, P(k,z) W(k x)^2} .
\label{eta-def}
\ee

The first approximation in  Eq.\eqref{eta-chain} means that, under some very generic conditions, the 
amplitude of the galaxy clustering ratio is the same in both real and redshift spaces. 
In the $\Lambda$CDM cosmology, this is exact at the linear level because the linear growing mode
$D_+(z)$, and its derivative ${\rm f}(z)=\partial \ln D_+/\partial\ln a$, do not depend on scale.
This is no longer true in modified gravity theories, which give rise to new scale dependences.
In addition, small-scale virial motions, which give rise to the fingers-of-god effect, also contribute
to the redshift-space power spectrum and to the clustering ratio. 
We shall discuss both effects in Sec.~\ref{sec:Redshift-space} and show that they do not 
impact the clustering ratio beyond the percent level.

The second approximation in Eq.\eqref{eta-chain} means that the amplitude of the galaxy clustering ratio 
is approximately identical to the amplitude of the analogous statistics for matter fluctuations.
This actually involves two properties, that the corrections due to nonlinearities of the biasing scheme
can be neglected, and that the scale dependence of bias coefficients does not sufficiently distort the 
shape of the power spectrum on the relevant scales to significantly modify the clustering ratio $\eta$.
Both effects will be addressed in turns in Sec.~\ref{sec:Non-linear-bias} and \ref{sec:Scale-dependent}.

\subsection{Expected accuracy and applicability} 
\label{sec:Expected-accuracy}

The $\eta$ formalism  was engeneered to ease the comparison  between data and theoretical
predictions.  From the observational perspective, the advantage of the galaxy clustering ratio rests on the  
simplicity and accuracy with which it can be extracted from redshift galaxy catalogs. Indeed, it provides
information on $P(k,z)$ without the need of reconstructing the galaxy power spectrum in Fourier space,  
nor the correlation function of galaxies, along with their covariance matrices;  one-point statistics such as counts in cells is all that is needed for its measurement. 
On the theoretical side, a distinctive feature of the $\eta$-statistic is the neat prediction of its 
amplitude which is virtually independent from any modelling assumption.
On large scales $x$ and $r$, where $P(k,z)$ is fairly described by a linear approximation,
and assuming standard gravity, {\it i.e.}  that  the linear growing mode  $D_+(z)$ does not depend on scale,
the amplitude of $\eta$ is not only independent from biasing and redshift-space distortion models, but also 
from linear growth rate of structures, cosmic time, and normalisation of the matter power spectrum. 
In other terms, there is no need to model and subsequently marginalise over these quantities, 
a procedure that  is known to degrade both the accuracy and the precision of cosmological probes.

Our approach follows from the observation, developed in this paper,  that at least one of these characteristic predictions breaks down  if modified gravity is responsible for the large scale distribution of matter.
Specifically, if  a scale dependent growing mode $D_+(k,z)$ is considered, $\eta$ is not a universal number anymore (at fixed scale $x$),  but becomes a  function of cosmic time,
as the time dependence no longer factors out in Eq.(\ref{eta-def}).
Clearly, this time dependent  signal is also expected in the $\Lambda$CDM cosmology if  $\eta$ is estimated on  scales where the linear approximation for the power spectrum breaks down.
To this purpose, note  that the highest  precision on  $\eta$  ($\sim 5\%$ from current redshift surveys of galaxies) can be achieved only on scales that are mildly non-linear.  For example, on the scale $x=16 h^{-1}$Mpc  used in this paper,  the  relative inaccuracy  induced on $\eta$ by adopting a linear power spectrum
instead of a non-linear one in Eq.\eqref{eta-def} is of order  $6\%$ for $n=2$ and $3\%$ for $n=3$, as found in FIG.~\ref{fig_etaxi_NL} below.
 Because of this reason, in the rest of this paper we will adopt a non-linear prescription for modeling the matter power spectrum.

What is also crucial for our discussion, is that the estimation of a characteristic  observable of the perturbation sector, such as the clustering ratio $\eta$,   only requires
the knowledge of the expansion rate of the universe, {\it i.e.}  prior information about the smooth sector of  the  theory only.
In other words, the measurement of $\eta$  does not  presuppose the premise to be tested,
{\it i.e.} the knowledge of a specific gravitational theory.
A prescription for converting redshifts into distances  is the only  ingredient needed for estimating  $\eta$ from redshift surveys data.
Since  a large class of  interesting modified gravity models predict  distance-redshift relations which are indistinguishable from that of the $\Lambda$CDM models,
the $\eta$ observable is such that we do not need to re-estimate it  in each of the  distinct gravitational scenarios we are testing.
From this remark follows the central argument of this paper: instead of assuming  a gravitational model and using $\eta$  to fix the expansion rate of the universe in that model,
as done by  BM14,  we here assume the expansion rate known from independent observations (specifically the Planck results \cite{planckXVI}) and use $\eta$ to distinguish
different gravitational models, specifically theories  where the Einstein-Hilbert action is supplemented by a general function  $F(R)$ of the Ricci scalar.

For this testing strategy being effective, the amplitude of the 
redshift-space galaxy clustering ratio $\eta^{\rm s}_{g}$, the quantity that can be
directly extracted from  galaxy surveys, should be predicted from the real space matter
power spectrum only (cfr. eq. \ref{eta-chain}) with an accuracy of $\sim 2\%$.  
This  level of accuracy is indeed enough to place
interesting constraints on possible deviations from the standard $\Lambda$CDM
scenario, notably on the family of $F(R)$ theories investigated in this paper.
That this precision is indeed achievable in the spatial and temporal regimes to which current data give access, 
is discussed in more details in Sec.~\ref{sec:Systematics}.

\section{The Clustering ratio and Screened $F(R)$ gravity}
\label{sec:f(R)-models}

\subsection{F(R) gravity}

As a template for the $F(R)$ gravity models, we choose  the bi-parametric form
\cite{Carroll:2003wy,Sotiriou:2008rp,Hu:2007nk}
\begin{equation}
F(R) = - 2 \Lambda c^2 - \frac{f_{R_0} c^2}{N} \frac{R_0^{N+1}}{R^N} ,
\label{fR-def}
\end{equation}
where  $f_{R_0}<0$ is a normalization factor and $N>0$. This Lagrangian  corresponds
to  the large-curvature regime of the model proposed in \cite{Hu:2007nk},
which is consistent with Solar-System and Milky-Way constraints thanks to the
chameleon mechanism, for $|f_{R_0}| \lessapprox 7 \times 10^{-7}$.  In the following
we will specialize our analysis to the cases $N=1$ and $2$.
The background dynamics agree with the reference
$\Lambda$CDM scenario with the same cosmological parameters.
The growth rate of density fluctuations, however,  is slightly modified. At the linear
level, this follows from the fact that the Newtonian potential $\Psi_{\rm N}$, or
Newton's constant, are effectively multiplied, in Fourier space,
by a scale dependent factor $1+\epsilon(k,z)$, where
\begin{equation}
\epsilon(k,z) = \frac{k^2}{3(a^2m^2+k^2)} ,
\label{epsilon-def}
\end{equation}
and
\begin{equation}
m^{-2} = 3 \frac{\dd^2 F}{\dd R^2} = -3 (N+1) f_{R_0}
\frac{c^2 R_0^{N+1}}{R^{N+2}} .
\label{m-def}
\end{equation}
On large scales, $k \ll a m$,  GR is recovered, whereas on small scales, within this
linear approximation, Newton's constant is larger by a factor $4/3$.
Stronger gravity implies that structure formation is favored and ultimately results
in a  matter  power spectrum amplitude which is larger than that characterizing the
$\Lambda$CDM model on mildly non-linear scales,
$k \sim 1 h {\rm Mpc}^{-1}$.
For smaller scales and high densities, non-linear effects are no longer negligible
and the chameleon mechanism ensures convergence  to GR.
As $|f_{R_0}|$ goes to zero, $m^2$ goes to infinity and  General Relativity is
recovered, hence the $\Lambda$CDM scenario, on all cosmological scales.
Hereafter, we consider as {\it reference} $\Lambda$CDM, the spatially-flat
six-parameter model  shown in column $1$ (\textit{Best fit}) of  Table 2
by \cite{planckXVI}.

The amplitude  of the clustering ratio expected in $F(R)$ gravity is computed
using the formalism described in \cite{BV2}. This combines one-loop
perturbation theory [that includes non-linear effects beyond the $\epsilon(k,z)$
factor, such as new quadratic and cubic vertices in the Euler
equation generated by the $F(R)$ theory] and a halo model [which takes into
account the non-linear impact of the $F(R)$ theory on the halo mass function
through the analysis of the modified spherical collapse]. This approach provides a
realistic estimate of the real-space matter density power spectrum, from large
scales to small scales, that is automatically consistent with one-loop
perturbation theory and agrees with numerical simulations up to their highest
available wave number, $k \lesssim 3 h$Mpc$^{-1}$ at $z=0$ \cite{BV2}.

\subsection{The clustering ratio from  SDSS data}
\label{sec:SDSS}

\begin{figure}
\begin{center}
\epsfxsize=8.8 cm \epsfysize=6.3 cm {\epsfbox{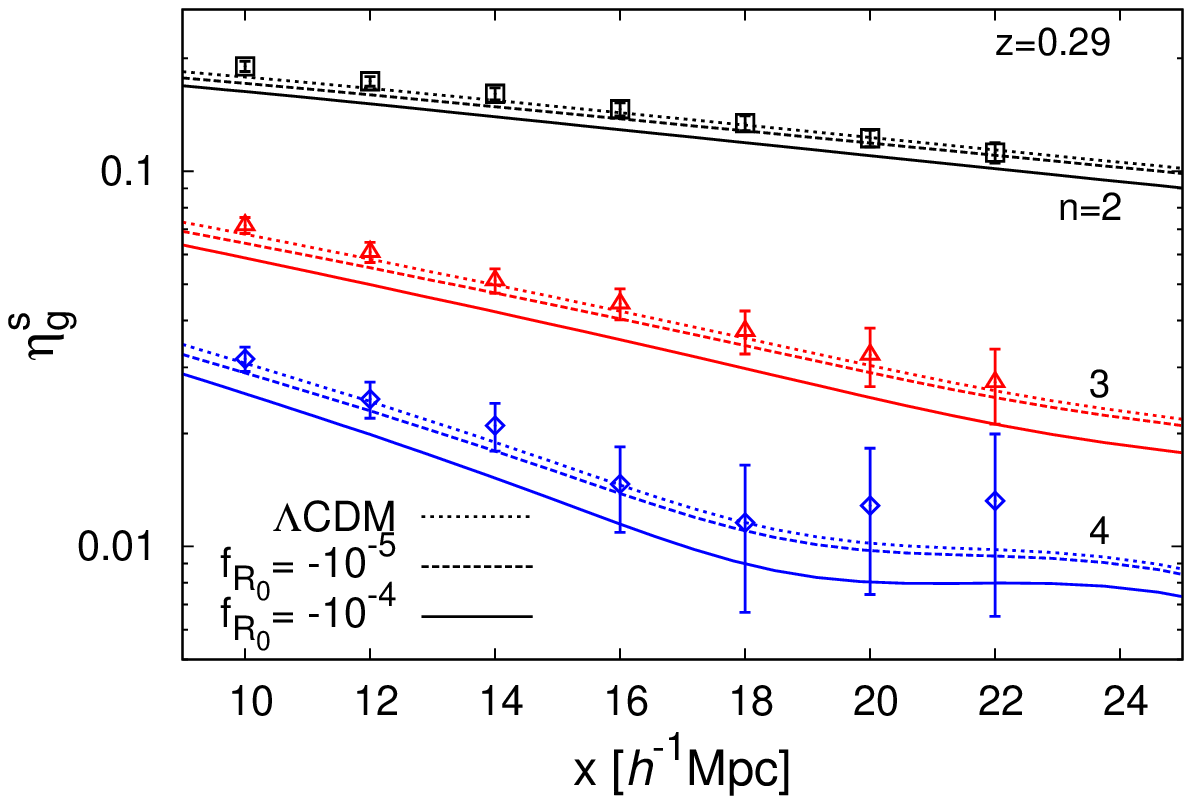}}\\
\epsfxsize=8.8 cm \epsfysize=6. cm {\epsfbox{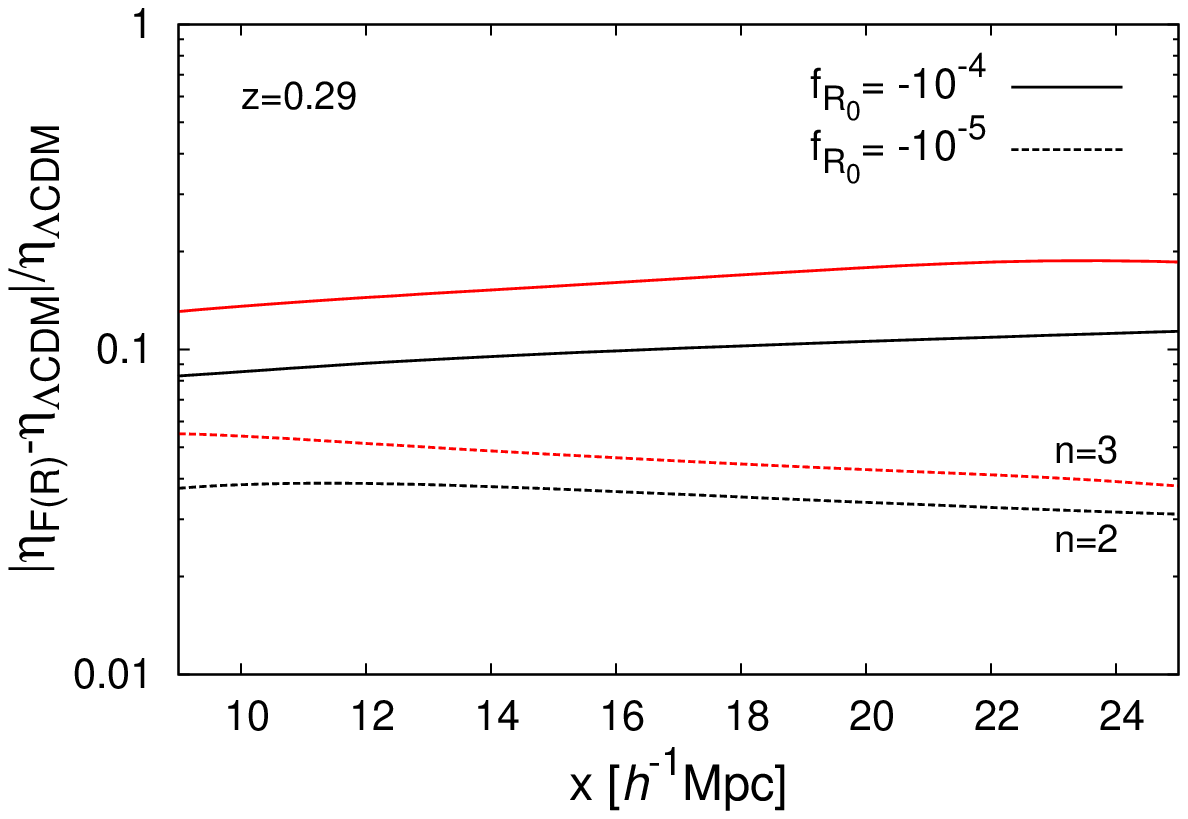}}
\end{center}
\caption{{\it Upper panel:} clustering ratio $\eta^{\rm s}_{g}(n,x)$ as a function of the
filtering scale $x$ for $n=2$ (squares), $n=3$ (triangles), and
$n=4$ (diamonds), at  $z=0.29$, the  mean redshift of the $s1$ sample.
The upper (dotted) line represents the scaling predicted in the reference
$\Lambda$CDM model (flat model with $\Omega_{m0}=0.3175$)
while the  middle (dashed) and lower (solid) lines correspond to  the $F(R)$ models
with exponent $N=1$ and normalisation parameters $f_{R_0}=-10^{-5}$
and $f_{R_0}=-10^{-4}$ respectively.
{\it Lower panel:} relative deviation of the $F(R)$ models from the $\Lambda$CDM
prediction.}
\label{fig_etaxi_fR_z0.29}
\end{figure}

We estimate the clustering ratio of the luminous red galaxy (LRG) sample extracted from the SDSS
data release 7 \cite{sdss7}  as well as from the data release 10 \cite{sdss10}. The first catalogue
($s1$)  covers the redshift interval $0.15 < z < 0.43$, has a contiguous sky area of $120 \times  45$
deg$^2$, and  comprises  $62,652$ LRG.
The second sample ($s2$),  extracted from the SDSS DR10 after removing all the objects in common  with $s1$,
extends over a deeper interval $0.3< z < 0.67$  but shallower (and not contiguous) field of view $\sim 3000$~deg$^2$.

The galaxy clustering ratio is estimated, assuming the redshift-distance conversion of the reference $\Lambda$CDM model (flat universe with $\Omega_{m0}=0.3175$),
as detailed in \cite{bm13b}.  Error bars are derived from a $30$ block-jackknife resampling of the $s1(/s2)$ data, excluding, each time, a sky area
of $12 \times 14$ deg$^2$(/$10 \times 10$ deg$^2$).  This specific scheme to estimate uncertainties  when  $\eta$ is estimated from    SDSS data was
shown to  give error bars  in excellent agreement ($\sim  8\%$ relative difference) with those
deduced from the analysis of the standard deviation displayed
by $40$ SDSS-like simulations (the LasDamas simulations \cite{ld}), which include, by
definition, the contribution from cosmic variance. This is suggestive of the fact that
$\eta$, being  defined as a ratio of equal order statistics,  and
thus containing the same stochastic source, is weakly sensitive to this systematic effect.

Results  for scales  $9 \leq x \leq 25 h^{-1}$Mpc and correlation indices $n=2,3,4$ are shown in  FIG.~\ref{fig_etaxi_fR_z0.29}.
Note that the lower limit on $x$ ensures  that quasilinear perturbation theory, the framework in which the $\eta$ formalism is developed, consistently  applies, while the upper limit on $n$
is set because measurements are progressively noisier  when the correlation length  $r=n x$ increases.
The scale $x=16h^{-1}$Mpc provides an optimal  trade-off that guarantees both  theoretical and observational accuracy,  and in the following we will only consider  the clustering ratio signal extracted on this scale.  Additional arguments that justifies  the choice of this filtering window will be provided  in the next section, where the overall systematic uncertainty affecting our analysis is presented and discussed.

A generic yet distinctive feature of  the matter power spectrum in $F(R)$ theories
is an excess of power  on  weakly  non-linear scales, $0.1 \lesssim k \lesssim 10 h$Mpc$^{-1}$,   with respect to the $\Lambda$CDM case.
On the scales considered here, $x>10 h^{-1}$Mpc,  we thus expect these theories to predict a smaller clustering ratio $\eta(n,x)$.
Indeed,  $\eta$ provides  a measure of the ratio of the power spectrum  at the characteristic scales $r=n x$ and $x$, that is,  on large scales, its amplitude is roughly  given by
\begin{equation}
 \eta(n,x)  \sim  \frac{   D_{+}^2(1/nx,z) \Delta^2_{L0}(1/nx)}{D_{+}^2(1/x,z) \Delta^2_{L0}(1/x)},
 \label{eq_app}
 \end{equation}
 where $\Delta^2_{L0}(k)= k^3P_{L0}(k)/2\pi^2$   is the initial linear power per logarithmic interval of $k$ and   $D_{+}(k,z)$ the linear growing mode.
The power suppression is effectively what is found in FIG.~\ref{fig_etaxi_fR_z0.29},  which  illustrates the scale dependence of $\eta$ in both  the reference $\Lambda$CDM and  $F(R)$ scenarios.
Note that the relative deviation between $F(R)$ and $\Lambda$CDM predictions is approximately constant, at least over the range of scales displayed, while its amplitude grows with the correlation index $n$. The uncertainty in the data, however,  grows even faster,   that is  the signal-to-noise ratio decreases as a function of $n$  thereby reducing the discriminatory power of the diagnostic at high $n$.

\begin{figure}
\begin{center}
\epsfxsize=8.8 cm \epsfysize=6.3 cm {\epsfbox{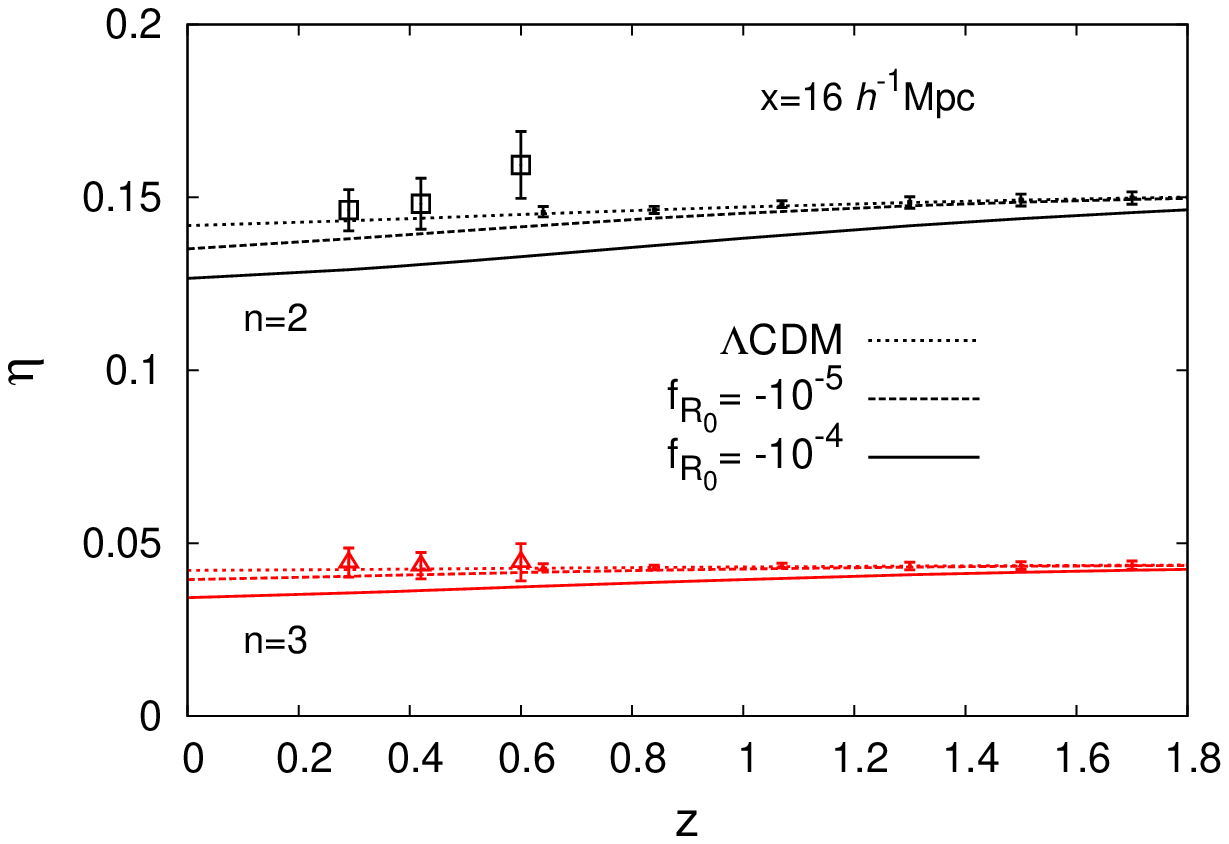}}\\
\epsfxsize=8.8 cm \epsfysize=6. cm {\epsfbox{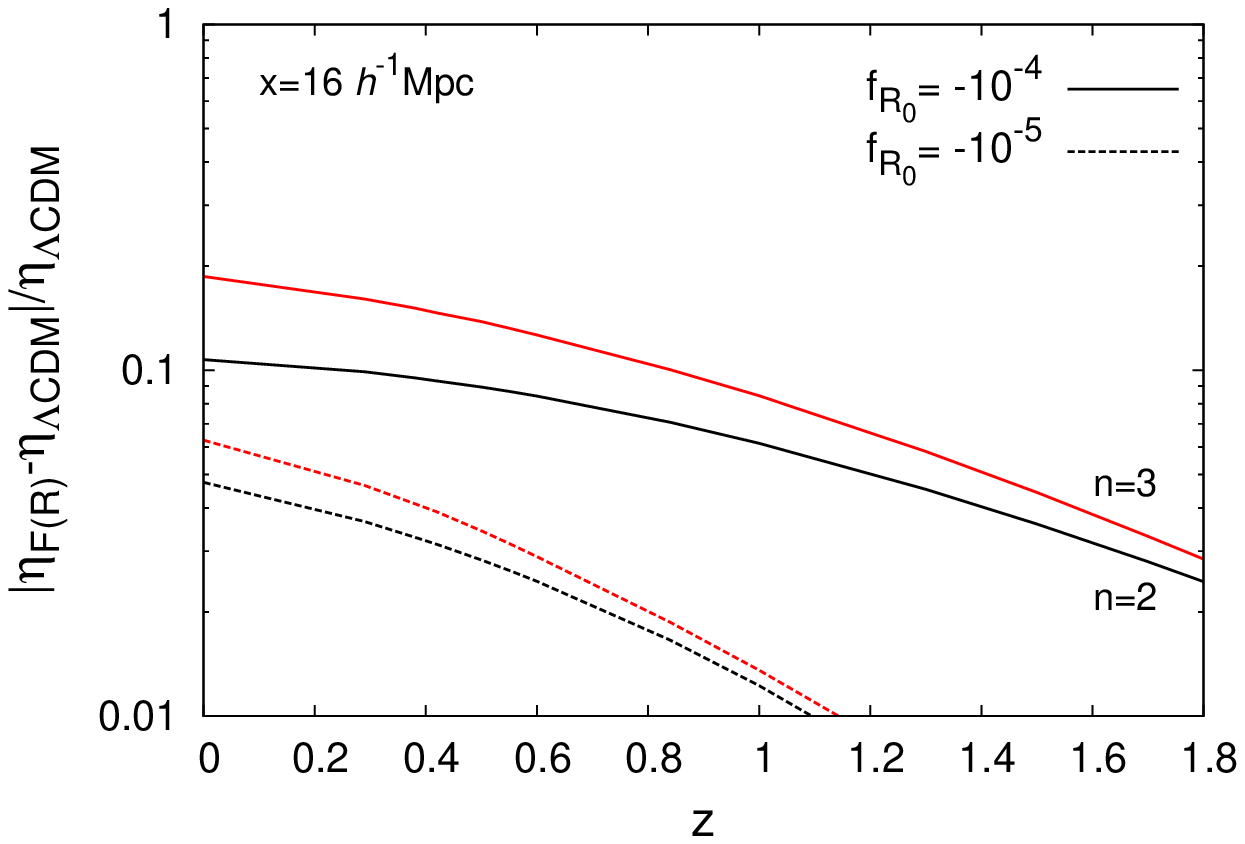}}
\end{center}
\caption{{\it Upper panel:} clustering ratio $\eta^{\rm s}_{g}(n,x,z)$ measured in three
different redshift intervals centered  at $z=0.29$ ($s1$ sample), $0.42$ and $0.60$ ($s2$ sample).
The redshift intervals in which the $s2$ sample is split are defined so that error bars are roughly equivalent to that estimated from  the $s1$ sample.
We show measurements obtained for the smoothing scale $x=16 h^{-1}$Mpc and for the correlation indices
$n=2$ (squares) and $n=3$ (triangles).
We also show the  amplitude of the clustering ratio predicted by the reference $\Lambda$CDM scenario (upper dotted lines) and
by the $F(R)$ models with exponent N=1 and with normalization parameters  $f_{R_0}=-10^{-5}$ (middle dashed lines) and  $f_{R_0}=-10^{-4}$ (lower solid lines).
We  give an example typical
of future surveys:
the small black error bars on the standard $\Lambda$CDM  curve (dotted) show forecasts for measurements
in bins of size $\Delta z \sim 0.2$ from an 15,000 sq. deg. survey of $7\times10^6$
galaxies, which closely matches what is expected from the Euclid mission \cite{euc}.
{\it Lower panel:} relative deviation of the $F(R)$ models from the $\Lambda$CDM prediction.}
\label{fig_etaxi_fR_x16}
\end{figure}

In FIG.~\ref{fig_etaxi_fR_x16} the amplitude of the clustering ratio  estimated at the three different
redshifts $z=0.29, 0.42$, and $0.60$ is shown (for  the typical quasi-linear scale   $x=16 h^{-1}$Mpc).
The clustering ratio signal of the  $s1+s2$ samples on the filtering and correlation scales  $x=16h^{-1}$Mpc and $n=2$ is recovered with  a relative inaccuracy  of $~3 \%$. This figure is indicative of the current performances of the $\eta$ test as a gravity probe. To better appreciate it, one can contrast this figure with the expected distortions in the clustering ratio signal
induced by a non standard growth of cosmic structures. This is done in  the lower panels of FIG.~ \ref{fig_etaxi_fR_z0.29} and \ref{fig_etaxi_fR_x16}, where we show the  relative difference (and the redshift scaling) between the amplitude of $\eta(n=2, x=16 h^{-1}$Mpc$, z=0.29)$ in the  $\Lambda$CDM and  $F(R)$ models.  For instance, the $\eta$ amplitude at $z=0.29$ in models with $f_{R_0} = - 10^{-4}(/ - 10^{-5})$ is nearly $10\%(/4\%)$ smaller than predicted by $\Lambda$CDM
\footnote{The deviation from $\Lambda$CDM is not ten times smaller for
$f_{R_0} = - 10^{-5}$ than for $f_{R_0} = - 10^{-4}$ because the transition scale
$2\pi/(am)$ of the kernel $\epsilon(k,z)$ of Eq.(\ref{epsilon-def}) becomes closer
to the scales $x=16 h^{-1}$Mpc and $r= nx$ that we probe.
Thus, at $z=0.29$ we have $am \simeq 0.038 (/0.119) h {\rm Mpc}^{-1}$
for $f_{R_0}=-10^{-4} (/-10^{-5})$, which gives $2\pi/(am) \simeq
166 (/53) h^{-1}$Mpc.}.
This is greater than the $2\%$ accuracy of our approximation (\ref{eta-chain}),
see Sec.~\ref{sec:clustering-ratio}, which shows that within our framework
we can constrain these $F(R)$ models down to $|f_{R_0}| \sim 10^{-5}$.

The sensitivity of the clustering ratio as a probe of the cosmological scenario
is further enhanced by the fact that
not only the amplitude of the signal is of relevance,  but also its different scaling as a function of redshift.  Indeed, while in a $\Lambda$CDM cosmology the amplitude of $\eta$ is expected to be almost  constant in time, in modified-gravity scenarios, such as $F(R)$ theories,
the scale dependence of the effective Newton's constant eventually results
in a substantial redshift dependence of the predictions for the amplitude of $\eta$.
In particular the discrepancy between the $F(R)$  and the $\Lambda$CDM predictions amplifies with time as they are indistinguishable at early cosmic epochs (see bottom panel of  FIG.~\ref{fig_etaxi_fR_x16}). A detection of a statistically significant redshift dependence of the   $\eta(n,x;z)$ signal  is therefore  a strong and unequivocal signature of deviations from the $\Lambda$CDM scenario.
Thus, in the $\Lambda$CDM case, from $z=1$ to $z=0$ the clustering ratio at
$x=16 h^{-1}$Mpc decreases by $3.7\%$, because of the nonlinear growth of the power
spectrum (which of course gives rise to a scale dependence as it introduces the nonlinear scale of matter clustering).
For $f_{R_0}=-10^{-4} (/ -10^{-5})$, it decreases by $8.4\% (/ 7.1\%)$. This greater decrease
at low redshift than for the $\Lambda$CDM case is due to the additional scale dependence associated with modified gravity, which now appears at both the linear and nonlinear levels.


\section{Systematics}
\label{sec:Systematics}

The next step is to make sure that residual systematic effects  do not compromise the effectiveness of the $\eta$ formalism in disentangling $F(R)$ models  from  the reference $\Lambda$CDM.   The accuracy of the  relation  $\eta^{\rm s}_{g}=\eta_{g}$ was tested  using various numerical simulations of the large scale structure of the universe  in a $\Lambda$CDM  model  by
\cite{bm13a,bm13b}. In particular, under blind test conditions,  the $\eta$ formalism was shown to  recover, in an unbiased way,  the value of the cosmological parameters used in the simulations.
Here, our purpose is to explore whether  the chain of approximations shown in Eq.\eqref{eta-chain} holds to percentage level precision also in the context of the $F(R)$ model of Eq.\eqref{fR-def}.
We also consider the impact of nonlinearities and baryonic effects on the matter power spectrum itself.

\subsection{Redshift-space distortions.}
\label{sec:Redshift-space}

\begin{figure}
\begin{center}
\epsfxsize=8.8 cm \epsfysize=6.5 cm {\epsfbox{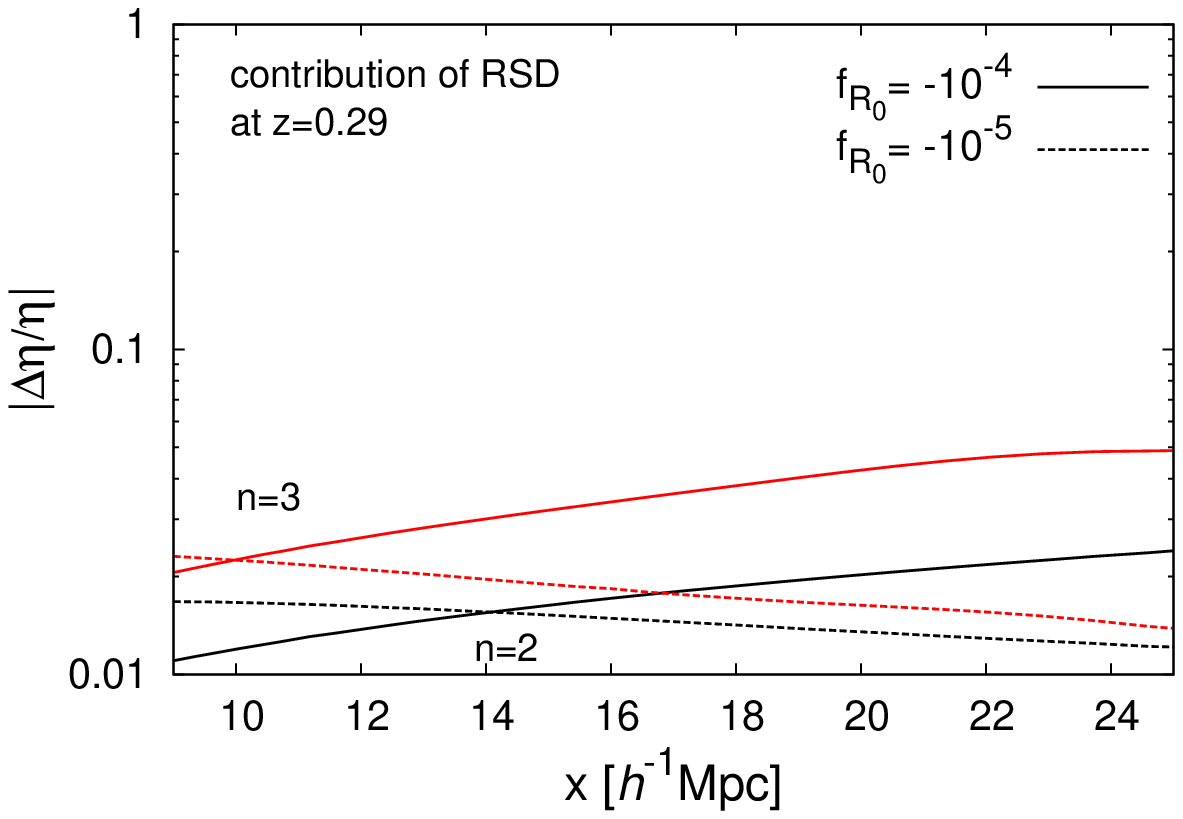}}\\
\epsfxsize=8.8 cm \epsfysize=6.5 cm {\epsfbox{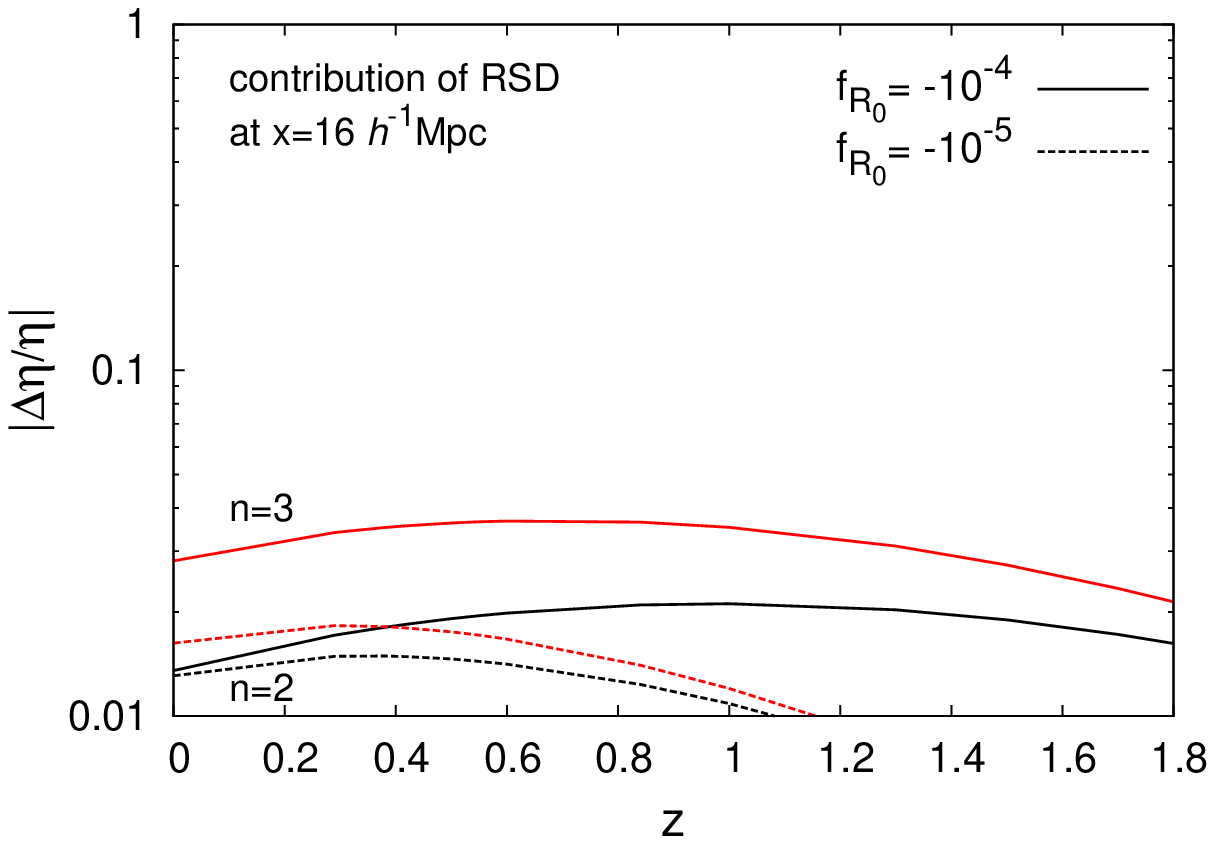}}\\
\end{center}
\caption{Impact of linear redshift-space distortions on the amplitude of the clustering
ratio, as a function of scale ({\it upper panel}) and redshift ({\it lower panel}). We show the
relative deviation between $\eta_g$ and $\eta_g^s$ [computed using the model in Eq.\eqref{beta}].}
\label{fig_etaxi_s}
\end{figure}

The first approximation in Eq.\eqref{eta-chain} is the use of the real-space clustering ratio
to estimate the observed redshift-space galaxy clustering ratio. Therefore, in this section we estimate
the impact of redshift space distortions (RSD).
In the $\Lambda$CDM cosmology, an interesting feature of the clustering ratio is that in the linear limit
it is insensitive to redshift space distortions, $\eta_{g}^{\rm s}=\eta_{g}$.  
This is because the linear Kaiser effect \cite{kai87} only multiplies the power spectrum by a factor 
$(1+\beta\mu^2)^2$ in redshift space, as in Eq.(\ref{beta}) below, and this scale-independent
factor cancels out in the ratio (\ref{etag-s-def}).
This simplicity is lost when we consider quasi-linear scales (where  non-linear motions are expected to 
contaminate the cosmological signal) or exotic models of gravity [where RSD might not factor out exactly in
Eq.\eqref{etag-s-def}].
This is the case in $F(R)$ scenarios, where the growth rate ${\rm f}(k,z)$ of linear matter fluctuations 
depends on the wave number $k$. 
As a consequence, a systematic bias results from neglecting the contribution of the RSD to the amplitude of 
the clustering ratio.
We evaluate quantitatively the amplitude of this bias by adopting the Kaiser model,
where we write the redshift-space power spectrum of galaxies [cf. Eq.\eqref{etag-s-def}] as
\be
P^{\rm s}_{g} (k,\mu) = b_1^2 (1+\beta \mu^2)^2 P(k) , \;\;\;
\beta(k,z) = \frac{{\rm f}(k,z)}{b_1} ,
\label{beta}
\ee
where $b_1$ is the linear galaxy bias in redshift space (although the matter real-space power
spectrum $P(k)$ includes non-linear corrections as explained in Sec.~\ref{sec:Nonlinear-spectrum}).
In the following we take $b_1=2$, a value well representing  the bias of
luminous red galaxies \cite{nuza}.
The relative error that results from  neglecting the linear RSD effect is shown in FIG.~\ref{fig_etaxi_s}.
As expected, the RSD correction is typically of order of the percent and smaller than that arising from 
neglecting to correct the power spectrum for non-linear effects, see
Sec.~\ref{sec:Nonlinear-spectrum} below.
It is also smaller than the deviation between the $\Lambda$CDM and $F(R)$ real-space
predictions for $\eta$, as the $k-$dependent growth rate ${\rm f}(k,z)$ is damped
by the cosine $\mu$ and the bias $b_1$.
Note that this error depends on the parameter $f_{R_0}$ and vanishes for $|f_{R_0}| \rightarrow 0$ as we converge to the $\Lambda$CDM cosmology.
In particular, for  $x=16 h^{-1}$Mpc, $z=0.29$ and  $n=2$, the relative contribution of the redshift-space distortions to the clustering signal is $1.5\%(/1.4\%)$ for $f_{R_0} = - 10^{-4}(/ - 10^{-5})$.

\begin{figure}
\begin{center}
\epsfxsize=8.8 cm \epsfysize=5. cm {\epsfbox{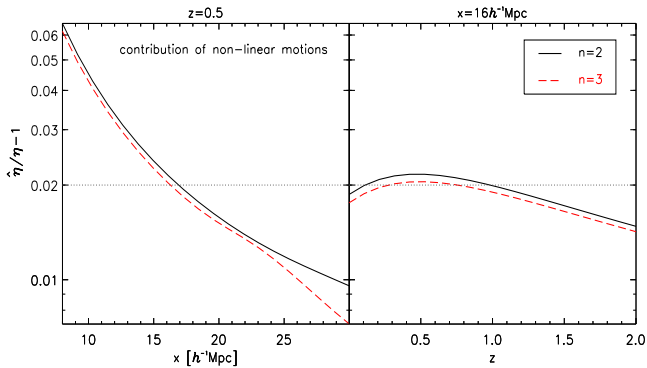}}\\
\end{center}
\caption{Impact of non-linear random motions on the amplitude of the clustering
ratio, as a function of scale ({\it left panel}) and redshift ({\it right panel}). We show the
relative deviation between the $\eta$ amplitude predicted before and after correcting the 
clustering ratio estimates with a Gaussian model of the galaxy velocity dispersion  
[cf. Eq.(9) of \cite{bm13b}].}
\label{fig_NLM}
\end{figure}

In addition to the corrections associated with large-scale coherent flows discussed above,
small-scale random motions of galaxies within virialized structures also contribute to the redshift-space
power spectrum, giving rise to the fingers-of-god effect. 
The leading order contribution of these small-scale effects to the amplitude of $\eta$ 
can be estimated by using Eq.(9) of \cite{bm13b}. This model assumes that a Gaussian kernel fairly  
describes the distribution of pairwise velocities along the line of sight, with a dispersion 
$\sigma_{12}=300 \rm{km/s}$.
The amplitude of the relative error induced by neglecting contributions from non-linear peculiar velocities
is shown in FIG.~\ref{fig_NLM}.  As expected, the systematic error decreases rapidly with the filtering scale 
$x$ (given the incoherent nature of small-scale non-linear motions), and appears to be almost insensitive 
to cosmic time (since motions induced by local gravity are detached from  the Hubble expansion). 
More importantly, on the scale relevant to our analysis ($x=16h^{-1}$ Mpc) the error is of order 
$\sim 2\%$ and comparable to the very same  precision ($\sim 1.6 \%$) with which phenomenological 
prescriptions available in the literature are effective in modelling these random motions (at least those 
simulated via  numerical experiments) \cite{bm13b}.
Because we consider modified gravity models that are very close to the $\Lambda$CDM cosmology,
we can expect these small-scale effects to keep the same order of magnitude.

Thus, the corrections to the clustering ratio due to large-scale coherent flows 
(more specifically, their scale dependence generated by modified gravity) and to small-scale
motions are of the same order of magnitude. 
Interestingly however, while the large-scale RSD tends to suppress the amplitude of the
clustering ratio on a given scale $x$, the non-linear random motions act in the opposite direction. 
The global resulting inaccuracy in the relation $\eta_{g}^{\rm s}=\eta_{g}$ is thus expected to be
smaller than $1\%$. We emphasize that this figure is much smaller than the precision with which 
the large-scale evolution of velocity fields is described by linear theory \cite{jbp}.   
Note, also, that these values are significantly smaller than the relative difference between the 
amplitude of $\eta$ predicted in  $F(R)$ models (with parameters $f_{R_0} = - 10^{-4}$ or 
$f_{R_0} = - 10^{-5}$) and the $\Lambda$CDM value. Therefore, the residual effects induced by the choice 
of not modelling redshift-space distortions  do not impair the ability of the clustering ratio to constrain
$f_{R_0}$ down to $|f_{R_0}| \simeq 10^{-5}$.

\subsection{Non-linear bias}
\label{sec:Non-linear-bias}

After redshift-space distortions, the second approximation in Eq.\eqref{eta-chain} is to neglect 
corrections due to galaxy biasing.
We first investigate in this section the accuracy of the statement that the clustering ratio amplitude 
is independent from the galaxy biasing function and its possible nonlinear character. 
As described in Refs.\cite{bm13a, bmmnras}, this result holds on those  scales $x$ where a local 
deterministic biasing scheme, $\delta_{{g},x}=\sum_i b_{i,x} \delta_{x}^i/i!$, fairly describes the relation 
between galaxy and matter density fields and the constraints $|b_{1,x}/b_{2,x}|>\sigma_{x}^2$ and
$1> | b_{2,x} \xi(r) |$ on the lower order biasing coefficients are both satisfied.
For example, on the scale $x=16h^{-1}$Mpc, the inaccuracy in the second approximation 
shown in Eq.\eqref{eta-chain} is $0.8(/0.6)$ \% for $n=2(/3)$ at $z=0.5$, and $0.8(/0.3)\%$ at $z=0(/1)$
for $n=2$.
These figures are computed by evaluating the contribution of higher-order, bias dependent corrections, 
using Eq.(42) of \cite{bmmnras}, under the assumption of a non-linear galaxy biasing scheme with 
$b_{1,\rx}=2$ and  $b_{2,\rx}=-0.2$, fairly representative of what is found  from the analysis of  red galaxy 
samples similar to those used in this paper  \cite{mbias1, mbias2, nuza}.

As a comparison, if, as usually done in the literature, one neglects higher order biasing contributions to the 
relation between the $rms$ of galaxy and matter fluctuations,
the precision of the approximated relation  $\sigma_{g,x} \approx b_1 \sigma_{\rm x}$  (on the same scale $x$ discussed above) 
is roughly $5$ times  poorer, being affected by a relative systematic error of nearly  $4\%$.

As we consider screened $F(R)$ models that are very close to the $\Lambda$CDM cosmology, we expect the galaxy biasing mechanisms to be essentially the same as in the standard cosmological scenario. 
Thus, the impact of non-linear biasing corrections should remain about $1\%$ or less. 
This is significantly smaller than the expected signal distortions induced by non-standard gravity
with $|f_{R_0}| \gtrsim 10^{-5}$, see  FIG.  ~\ref{fig_etaxi_fR_x16}, and below the accuracy of $2\%$ 
that we aim at in this paper.
Therefore, we can neglect these nonlinear biasing corrections for our purposes.
This also simplifies the analysis as it avoids resorting to a more refined, bias-dependent, theoretical 
prediction for the $\eta$ amplitude.

\subsection{Scale-dependent bias}
\label{sec:Scale-dependent}

We have seen in the previous section that nonlinear biasing does not give rise
to significant corrections on the large scales that we consider in this paper.
However, even within a linear bias model, another source of systematics
due to the bias arises from the scale dependence of the galaxy bias, which
can distort the shape of the correlation function and mimic the scale-dependent
growth associated with a modified-gravity scenario.

\subsubsection{Sensitivity to scale-dependent biasing}
\label{sec:Sensitivity-to-scale-dependent biasing}

\begin{figure}
\includegraphics[width=75mm,angle=0]{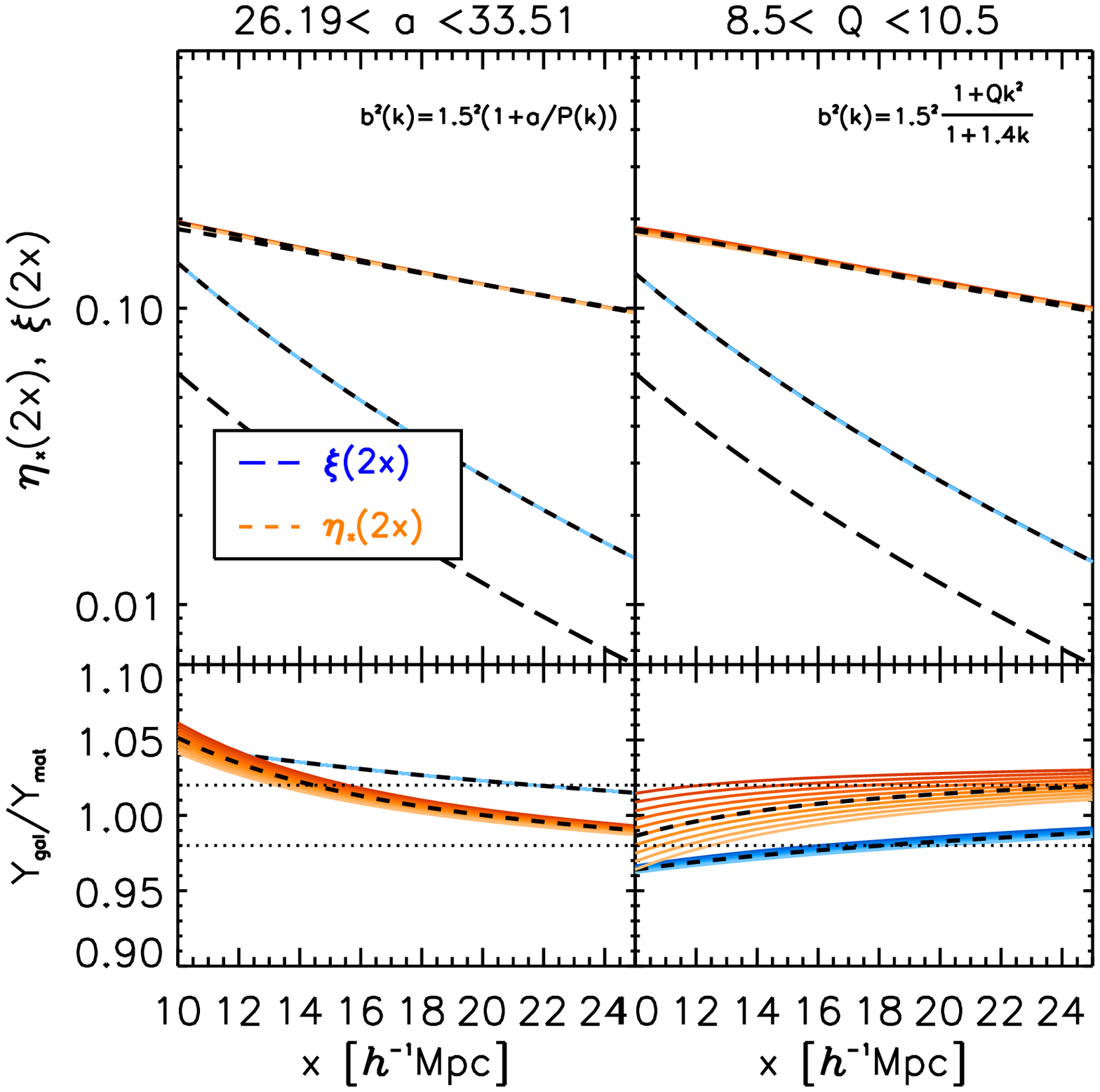}
\includegraphics[width=75mm,angle=0]{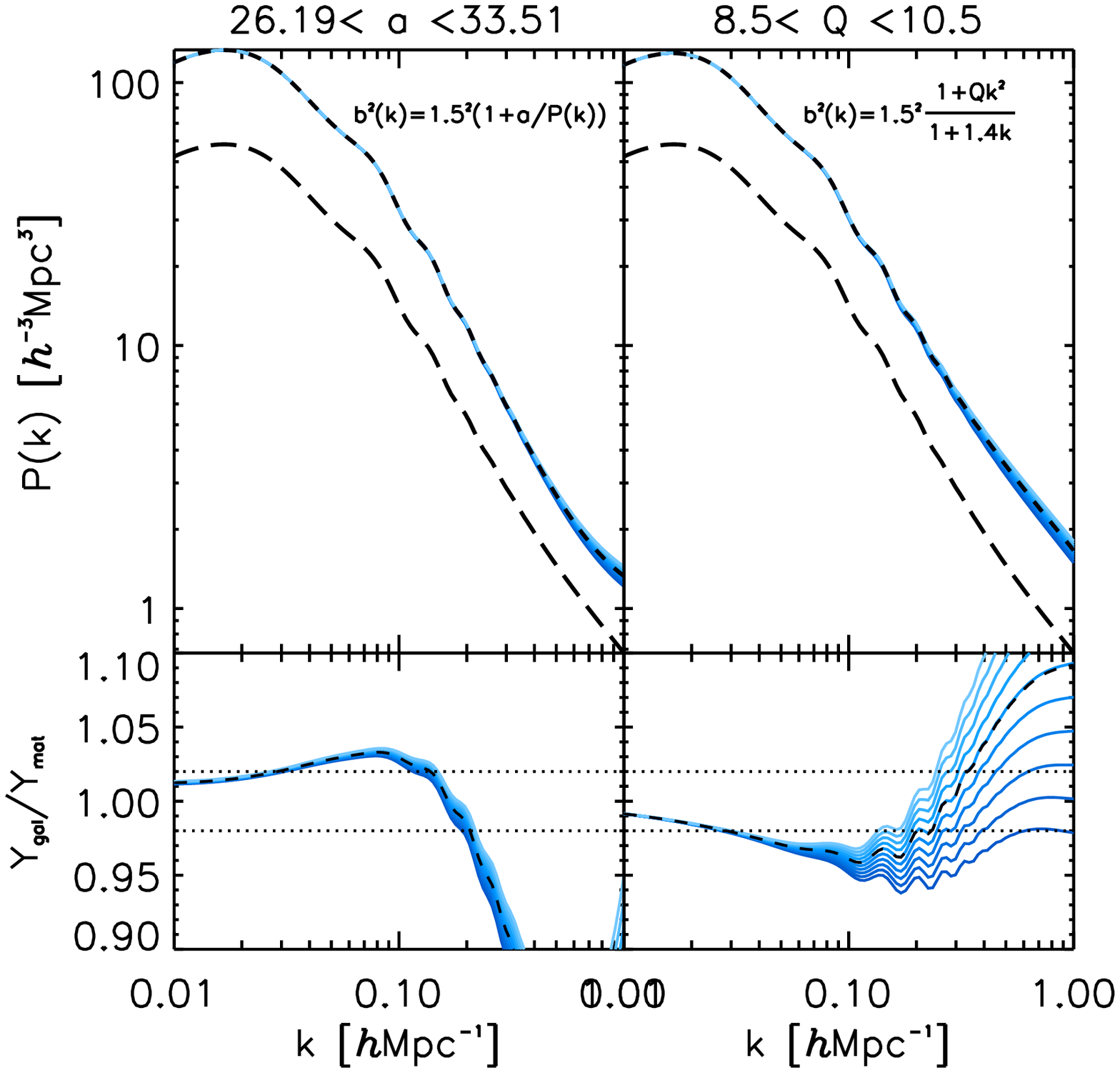}
\caption{Impact of  scale-dependent contributions of the biasing models, here the P- and Q-models, on the ratio between galactic and matter quantities $Y$ where $Y$ can be either the power spectrum $P(k)$, the 2-point correlation $\xi(x)$ or
 the clustering ratio $\eta(2,x)$. The characteristic parameters of the biasing models ($a$ and $Q$) are varied within their
 $95\%$ confidence interval determined by \cite{cresswell2008}.
 In all cases, the $\eta$ dependence on bias is less than two percent.}
\label{fig_Peta}
\end{figure}

Since  the clustering ratio is engineered to extract cosmological information encoded in the galaxy distribution on a given filtering scale $x$, this statistics is by construction  independent from any possible (real-space) scale dependence $b(x)$ of the biasing function
on smaller scales. Thus, the $\eta$ formalism is built upon the hypothesis that biasing is local,  i.e. by explicitly neglecting the possibility that galaxy and matter  power spectra are related  by scale dependent operators  in Fourier-space [$P_{g} (k) = b^2 (k) P(k)$] on larger scales $r$ beyond the filtering scale $x$.
However, some degree of  biasing depending on the wave number $k$ is naturally expected on cosmological scales.
Fortunately, it is unlikely that neglecting this effect on large scales $r$  induces  an appreciable systematic error  in the predicted  amplitude of $\eta_{g}$.
Indeed, tests performed on LasDamas numerical simulations of the large scale structure of the universe also confirmed, independently,  that possible systematic effects induced by a non local,  $k$-dependent galaxy bias can be safely neglected  on the scales explored in our analysis  \cite{bm13b}. 

This is well illustrated by the following example in which we consider the Q-model  $b^2(k) = b_1^2(1 + Qk^2)/(1 + Ak)P_L(k)/P(k)$ \cite{cole2005} with parameters  $A = 1.4$ and $Q = 9.6$ \cite{cresswell2008}. The relative variation of the squared bias, $\Delta b^2/b^2$, is as high as $8\%$ in the interval $0.01 < k < 1 h$/Mpc, but results in  $\eta$ changing by only $\sim 0.8\%$ on the relevant scales $x=16h^{-1}$Mpc and $r=2 x$. 

This substantial independence of the clustering ratio on scale-dependent
biasing is also illustrated in FIG.~\ref{fig_Peta}, where the ratios
$Y_{g}/Y_{\rm matter}$ are represented for both the Q-model and the
P-model, defined by $b^2(k)=(b_{1}^2 P_L(k)+a)/P(k)$, which, according to \cite{Smithetal}, has a more solid grounding in physics than the Q-model (the parameter $a$ corresponds to a shot-noise
contribution that can arise if galaxies Poisson sample the matter density field). The quantities $Y$ are either the power spectrum, the correlation function $\xi$ or the clustering ratio $\eta$.
In each case, the quantity $Y_{\rm matter}$ is multiplied by a constant bias $b_1=1.5$,
which actually cancels out in the case of $\eta$.
The width of the curves shows the impact of the variation of the bias parameters $a$ and $Q$ within
their $2\sigma$ confidence range \cite{cresswell2008}.
As can be seen in the figure, the scale dependence of the bias predicted by these models does not modify
$\eta_g$ by more than two percents, at $x=16 h^{-1}$Mpc, whereas it has a significantly greater effect on 
the power spectrum and the correlation function.
Moreover, varying the parameters of these bias models within $2\sigma$ intervals does not further
modify $\eta$ beyond $2\%$.
This is particularly important for the applications to modified gravity, as the knowledge of the biasing 
function and the range of the biasing parameters could be affected as compared with the $\Lambda$-CDM cosmology. These results show that the dependence of the clustering ratio on the scale dependence of the bias, through both the biasing parameters and the functional form of the biasing model, is within the required accuracy in order to derive sensible bounds on modified gravity.

\subsubsection{Comparison with marginalizing analysis based on the power spectrum}
\label{sec:Comparison}

\begin{figure}
\includegraphics[width=80mm,angle=0]{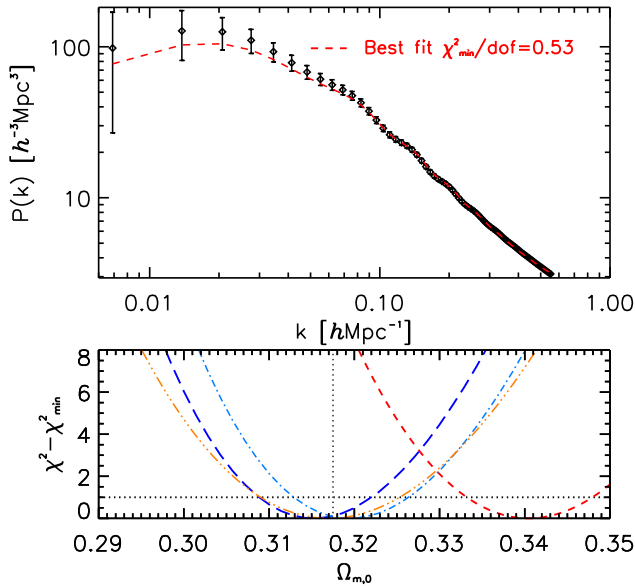}
\caption{{\it Upper panel:} galaxy power spectrum (diamonds) simulated using the {\it reference} $\Lambda$CDM model ($\Omega_{m0}=0.3175$) and the Q-model for describing scale-dependent galaxy biasing ($b_1=1.5$, $Q=9.6h^{-2}$Mpc$^2$ and $A=1.4h^{-1}$Mpc).  Error bars roughly correspond to what is expected in a survey like  BOSS \cite{sdss10}. The thick red dashed line shows the best fit that is obtained if we analyse this
data by the P-model (i.e., a ``wrong'' bias model), marginalising over $0.9<b_1<3$ and $10<a<60$
with flat priors. The minimum value of the $\chi^2$ statistic (with $\nu=100-3$ degrees of freedom) is indicated in the inset and  suggests that this best fitting model cannot be rejected on statistical grounds.
{\it Lower panel:} likelihood constraints on $\Omega_{m0}$. 
The vertical grey dotted line indicates the input value of the mass density parameter $\Omega_{m0}$. 
As in the upper panel, the thick red dashed line shows the result obtained from the power spectrum
analysis by using the ``wrong'' P-model, which leads to a significant overestimation bias.
The thin red dot-dot-dot-dashed line shows the 1D likelihood profile obtained from the power spectrum analysis after marginalising over the biasing parameters of the ``true'' Q-model itself, which gives
an unbiased result. 
These results are compared with those obtained by adopting the clustering ratio (with $x=16h^{-1}$Mpc and $n=2$) as observable in the likelihood analysis, without implementing any marginalisation scheme.
The corresponding likelihood profile is shown by the thick blue long dashed line. In addition, we also display with the blue thin dot-dashed line the likelihood profile obtained from the  $\eta$-test when the data are generated with the P-model.}
\label{fig_pk}
\end{figure}

A more traditional approach, especially when one uses the power spectrum itself as a probe of cosmology rather than the clustering ratio, is to marginalize over the nuisance parameters of the biasing model. Within a specific bias model, this allows one to take into account the possible change of the bias parameters associated with $F(R)$ scenarios. This approach, however, overlooks the possibility that the shape of the biasing function in $F(R)$ scenarios might be poorly described by the reference biasing model, and therefore that systematic errors might be introduced when marginalising over an improper function.
 
To substantiate this argument,  we simulate the matter power spectrum in an ideal cubic galaxy survey whose side is $\sim 900 h^{-1}$Mpc (which roughly corresponds to the volume surveyed by BOSS). We assume  Planck values for  the relevant $P(k)$ parameters, in particular $\Omega_{m0}=0.3175$. 
The galaxy power spectrum is then simulated by biasing the matter power spectrum with the Q-model.
We then try to retrieve  the input value of the  matter density parameter by means of  a Fisher analysis that uses as observable the shape of the power spectrum and which is run by marginalising over either the nuisance parameters $b_1$, $Q$  of the Q-model (i.e. we analyse the data with the ``true'' biasing scheme) or the parameters $b_1$, $a$ of the P-model (i.e. we analyse the data with a ``wrong'' biasing scheme).
In the bottom panel of FIG.~\ref{fig_pk}, we show that constraints on $\Omega_{m0}$ obtained after marginalising over the correct biasing model are both accurate, the true value of $\Omega_{m0}$ is within the $1\sigma$ interval, and precise, although the relative imprecision of the measurement ($\sim 2.5\%$) is  one order of magnitude larger than the imprecision ($\sim 0.2\%$ ) that could be attained if the values of the biasing  parameters were perfectly known.  
On the other hand, if the power spectrum analysis is carried out by marginalising over the parameters of the ``wrong'' biasing model, the inferred value of $\Omega_m$ is systematically larger, and its $1\sigma$ error bar does not  bracket anymore the true value. 

For comparison, we also show in the lower panel of FIG.~\ref{fig_pk} the likelihood profiles obtained 
from the $\eta$-test, without any marginalisation, when the galaxy power spectrum is computed with
either the Q-model or the P-model.
As expected, the $\eta$-test shows a small sensitivity to the scale-dependence of the bias,
as seen from the fact that the best-fit $\Omega_{m0}$ is not exactly the same if we use the Q-model or
the P-model. However, in agreement with FIG.~\ref{fig_pk}, in both cases the true value of
$\Omega_{m0}$ is within the $1\sigma$ error bar of the likelihood.
This again highlights the virtues of a  probe that, being by construction less sensitive to scale-dependent 
bias, also minimises possible systematics induced by  an improper marginalisation procedure.

In summary, we have shown that the clustering ratio can be used at the two percent level even though different biasing models with different ranges of parameters have been implemented. Moreover, the clustering ratio does not require to marginalise over poorly known biasing functions and parameters in modified gravity. Indeed, the required two percent accuracy can be attained without any marginalisation.

\subsection{Nonlinear matter power spectrum}
\label{sec:Nonlinear-spectrum}

\begin{figure}
\begin{center}
\epsfxsize=8.8 cm \epsfysize=6.5 cm {\epsfbox{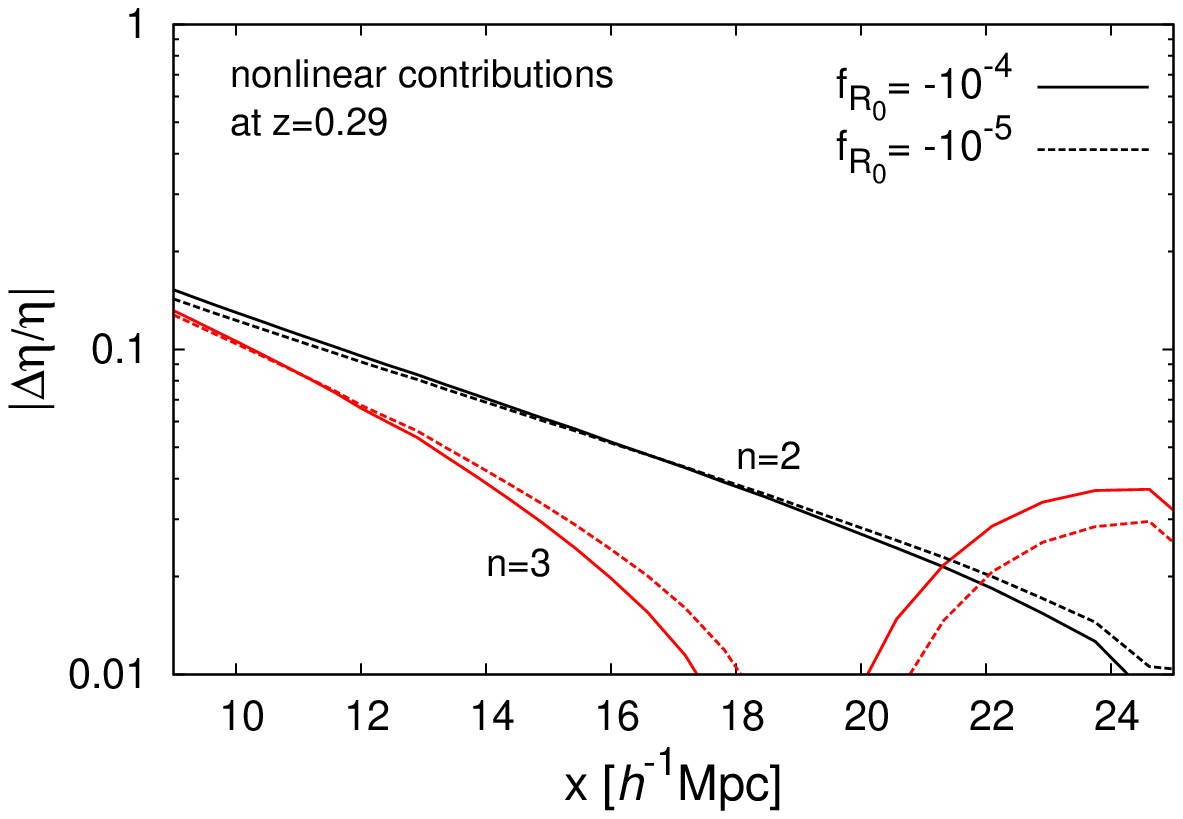}}\\
\epsfxsize=8.8 cm \epsfysize=6.5 cm {\epsfbox{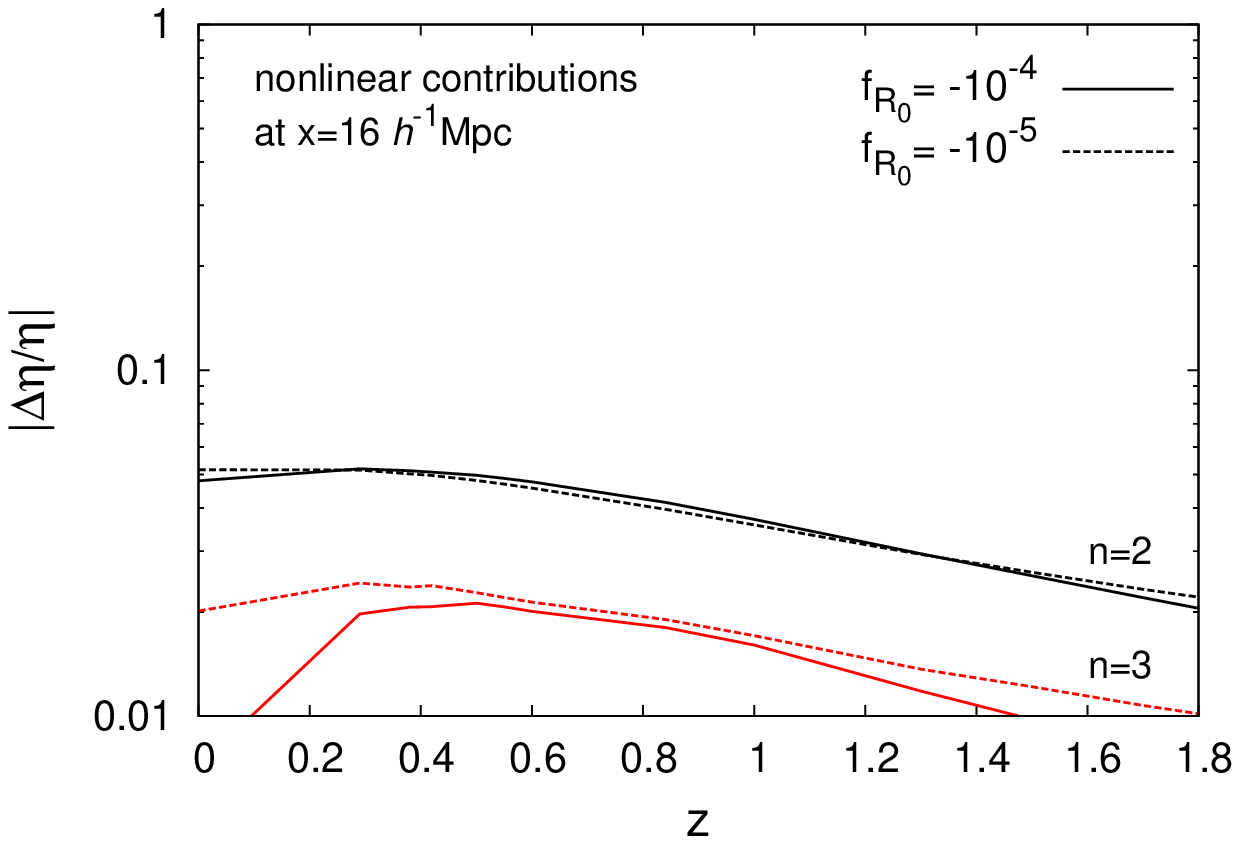}}\\
\end{center}
\caption{Impact of  non-linear contributions  the power spectrum of matter on the amplitude of the clustering
ratio, as a function of scale ({\it upper panel}) and  redshift ({\it lower panel}). We show the
relative deviation between $\eta$ predictions obtained using  the  linear theory and  the
non-linear model described in \cite{BV2}.}
\label{fig_etaxi_NL}
\end{figure}

We have seen in the previous section that the galaxy clustering ratio can be estimated from the matter
real-space clustering ratio. This greatly simplifies the analysis, but we still need to obtain sufficiently
accurate predictions for the matter power spectrum itself.
Indeed, the space volume occupied by the $s1$ and $s2$ samples forces us to evaluate the clustering ratio
on  scales $x$ that are not large enough for linear perturbation theory to be consistently applied.
On these quasi-linear scales, where high-order, model-dependent corrections to the power spectrum of 
matter cannot be in principle neglected, the density fluctuations are no longer separable functions of 
cosmic time $t$ and scale $k$. 
Therefore, the $\eta$-statistics acquires a characteristic dependence on the redshift 
(see FIG.~\ref{fig_etaxi_fR_x16}).
In FIG. ~\ref{fig_etaxi_NL}, we show the impact of including non-linear contributions to the power  spectrum 
$P(k)$ when calculating  the amplitude of the clustering ratio $\eta^{\rm s}_{g}$.
Specifically, we show the relative difference  between the $\eta$ amplitudes calculated by inserting 
into Eq.\eqref{eta-def} either the linear matter power spectrum $P_L(k)$ or the non-linear  model of 
\cite{BV2} (a real-space matter power spectrum that is exact up to order $P_L^2$ in perturbation theory).  
At the characteristic scale $x=16 h^{-1}$Mpc,
non-linear contributions to $\eta$ modify linear expectations  by only a few percents ($\simeq 6\%$ at 
$z=0.29$ for $n=2$, and much less for the larger correlation scale $n=3$).
Given that we aim at $2\%$ accuracy, and since this inaccuracy is larger than  the relative error with which 
the $\eta$ statistics can be measured  from current data ($\sim 3\%$),  in what follows we will incorporate in 
our analysis these non-linear corrections to the power spectrum of matter.
To this purpose, we use the analytical model described in \cite{VNT13,BV2}, which is exact
up to one-loop order of perturbation theory,  and matches $\Lambda$CDM numerical
simulations up to $2\%$ at $r \geq 16 h^{-1}$Mpc for $\xi(r)$.
In particular, as shown in \cite{V13-acc}, uncertainties due to non-perturbative small-scale
effects, such as a change of up to $10\%$ of the halo mass function or of the
mass-concentration relation (or using different published fits) for the underlying halo model,
only change $\xi(r)$ at $r \geq 16 h^{-1}$ Mpc by less than $1\%$.
Uncertainties in modeling the non-linear power spectrum of matter are thus expected
to affect in a negligible way the clustering ratio statistics, at least on scales
$x \geq 16 h^{-1}$ Mpc.

\subsection{Baryonic effects.}
\label{sec:Baryonic}

On small cosmic scales, the matter density power spectrum is also sensitive to the physics of
baryons and to galaxy formation processes, such as AGN feedback.
However, from FIG.~5 of \cite{Puchwein}, numerical simulations suggest that these
effects are small on large scales, $k < 1 h {\rm Mpc}^{-1}$ at $z=0$, and only reach the level
of the modification associated with an $F(R)$ model with $f_{R_0}=-10^{-5}$
at $k \gtrsim 7 h {\rm Mpc}^{-1}$. In configuration space, this leads to a damping of the density fluctuations on scales
smaller than $6 h^{-1}$Mpc.
Therefore, by considering larger scales, above $16 h^{-1}$Mpc at $z \geq 0.29$,
and restricting our analysis to $| f_{R_0} | \geq 10^{-5}$, we can safely neglect these effects.
Interestingly, the $\eta$ statistics is expected to be less sensitive to these local
effects than the power spectrum $P(k)$.
Indeed, the clustering ratio $\eta=\xi_x(r)/\sigma_x^2$, being a  statistics defined in configuration space and smoothed over scale $x$, should
be insensitive to redistributions of matter on smaller scales (whereas local motions typically
lead to power law tails $\propto k^4$ for power spectra \cite{Peebles74}).

\subsection{Robustness of the $\eta$ probe.}
\label{sec:Robustness}

In conclusion, within the regime of quasi-linear filtering scales ($x=16h^{-1}$Mpc) and moderate redshifts ($z<0.67$)  under investigation,
the $\eta$ statistics allows us to tell apart  standard and non-standard models of gravity  at the two percent level even without  the need of correcting predictions with  models for  non-linear bias, non-linear galaxy motions, or linear bulk flows. However, the predicted amplitude of  $\eta$
is still sensitive, in the regimes under investigations,   to the  modelling of the non-linear power spectrum.  Interestingly, however,
FIG.~\ref{fig_etaxi_NL}  shows that the Euclid space mission
will soon probe volumes of space large enough to make the estimation of $\eta$ independent also from high order corrections of the matter power-spectrum. Indeed,  once $\eta$ is estimated for $n=2$ on scales $x>25h^{-1}$Mpc,  these model-dependent corrections can be neglected to better than $1\%$ accuracy.

One last point deserves mention. While for  $f_{R_0}=-10^{-4}(/-10^{-5})$  the relative deviation between  $F(R)$ and $\Lambda$CDM predictions for the amplitude of $\eta$ is larger than the precision with which the clustering ratio is currently measured ($3\%$), this deviation
becomes smaller than $\sim 0.5 \%$ for $|f_{R_0}| < 10^{-6}$.
Therefore, it is unlikely that constraints on $F(R)$ models with $|f_{R_0}|< 10^{-6}$ be free  of systematics unless the neglected effects or the residual systematics affect in  the same way both the $F(R)$   and $\Lambda$CDM models.

\begin{figure}
\begin{center}
\epsfxsize=8.8 cm \epsfysize=5.6 cm {\epsfbox{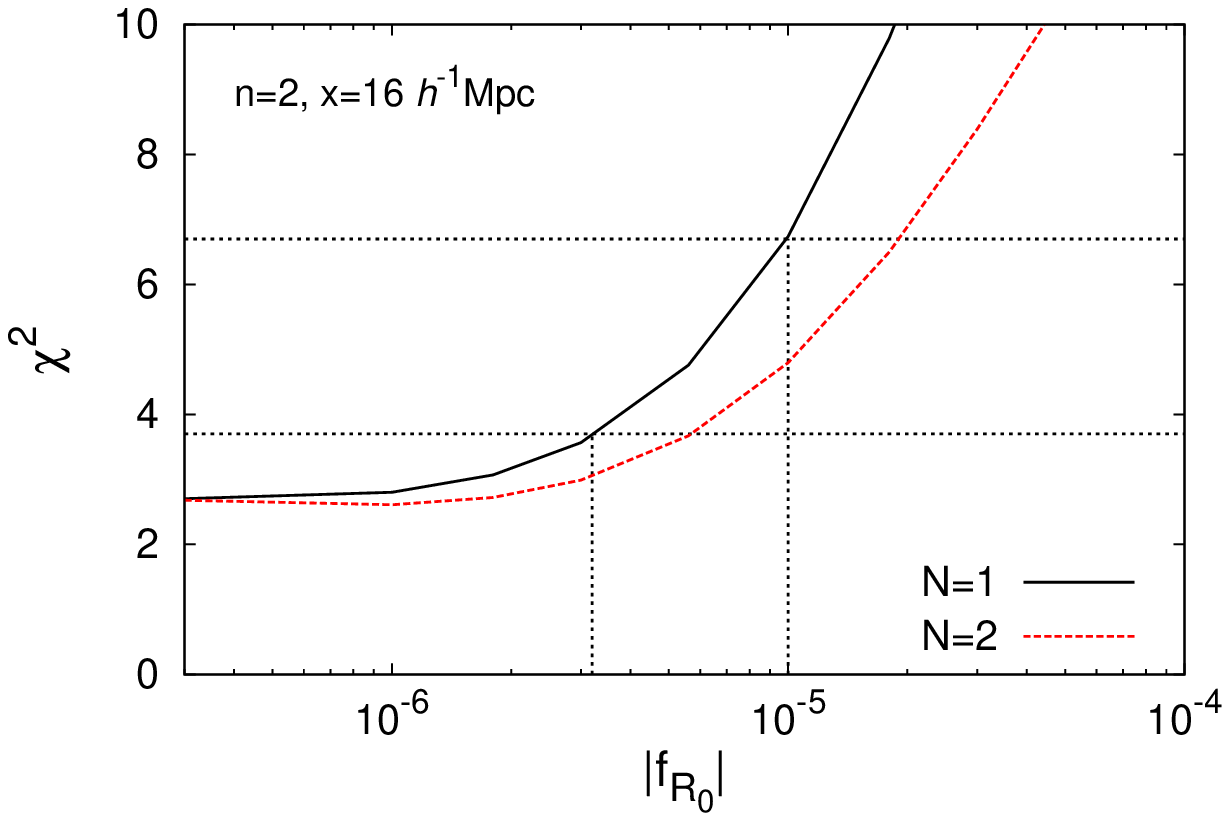}}\\
\epsfxsize=8.8 cm \epsfysize=5.6 cm {\epsfbox{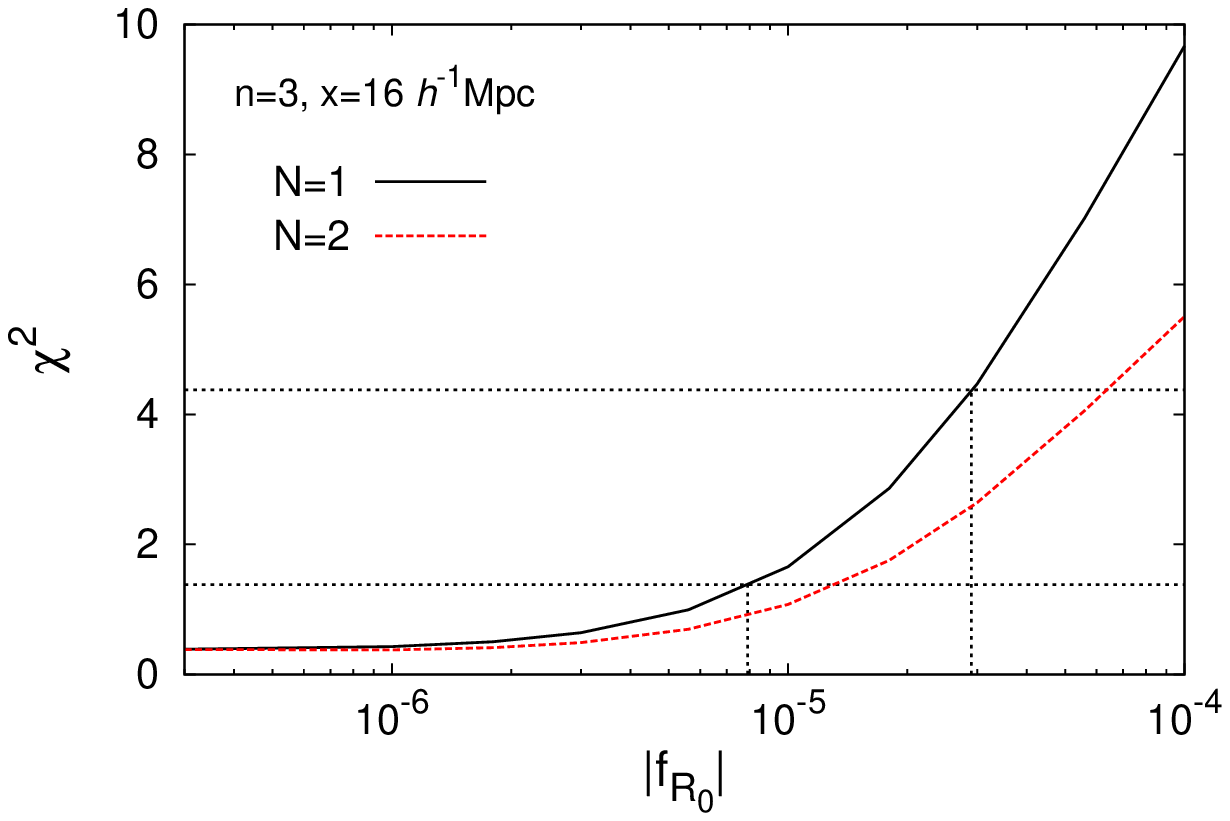}}
\end{center}
\caption{$\chi^2$ profile from the least square analysis of the clustering ratio data  $\eta(n,x)$ measured from the SDSS samples $s1$ and $s2$ for  $x=16 h^{-1}$Mpc and in three different redshift bins at $z=0.29, 0.42$, and $0.60$.
We show  results for the correlation indices $n=2$ ({\it upper panel}) and $n=3$
({\it lower panel}).
We consider  $F(R)$ models with $N=1$ (upper solid lines) and $N=2$ (lower dashed lines).
The horizontal dotted lines are the $68\%$ and $95\%$ confidence contours.}
\label{fig_chi2-all_z}
\end{figure}

\section{Constraints on $F(R)$ models}
\label{sec:constraints}

We quantify the confidence level with which current data reject an $F(R)$ gravitational scenario  by means of the standard $\chi^2$ statistic.
We do this by exploring two complementary scenarios. We  first consider $F(R)$  models with exponents $N=1$ or $N=2$ and we assume that
the background field value $|f_{R_0}|$   is  the only free fitting parameter. Therefore, in what we call hereafter {\it scenario 1},  we assume that the background expansion is exactly
described by  the reference $\Lambda$CDM model. By this choice we want to mimic the situation in which background data have infinite precision and the discriminatory power
on modified gravity parameter is provided  only by perturbed sector data.  This scenario also allows us to highlight, neatly,  the specific virtues of the clustering ratio as a diagnostic of gravity.
 In {\it scenario 2} we take into account the uncertainty with which the background expansion history is presently known by
allowing  for an additional fitting parameter,  the matter density parameter $\Omega_{m0}$, which is not known to better than $6\%$ ($68\% \,c.l.$  from
Planck data).  We  still assume, however,  that the expansion of the background is fairly described in terms of a flat $\Lambda$CDM model and that the remaining parameters to which $\eta$ is sensitive,  $n_s$, $\Omega_b h^2$, and $H_0$, are  fixed to their Planck value [their uncertainty ($0.7\%, 1.2\%, 1.8\%$ respectively) being  negligible with respect to that of the matter density parameter].
An additional parametric dependence, specifically
on the $rms$ of matter fluctuations $\sigma_8$, naturally arises as a consequence of   estimating  $\eta$ on quasi-linear scales, i.e. on scales where this parameter controls the  shape of the non-linear power spectrum of matter. On the scales explored in this paper,
the functional dependence of $\eta$ on $\sigma_8$  is however extremely weak.  Indeed, although the relative uncertainty  $\Delta \sigma_8/ \sigma_8$  is rather large ($ \sim  3\%$) if compared to the precision achieved by Planck on other parameters, varying $\sigma_8$ within the Planck $99.7\%$ confidence interval  only results in $\eta$ changing by  $0.2\%$ at most  (a figure that should be compared, for example, with the relative change with respect to the best fitting value  $\delta \eta/\eta\sim -14\%(+18\%)$ when $\Omega_m$ is estimated at the upper(/lower) extrema of  the Planck $99.7\%$ confidence interval).

Because the signals at different scales $x$ are correlated, we only analyse, in both scenarios,  the galaxy field filtered
on the scale $x=16 h^{-1}$Mpc,  a trade-off between the precision of measurements (worsening as $x$ increases) and of theory, {\it i.e.} of  Eq.(\ref{eta-def})  (worsening as  $x$ decreases).
We make separate analyses for the correlation indices $n=2$ and $3$,  hereafter called respectively {\it reference} analysis and {\it control} analysis. Note that  the covariance matrix is diagonal,
since  the $\eta$ measurements  in the three different  redshift bins can be considered as independent estimates.

The  resulting  1D Log-likelihood profiles  obtained from the analysis of the clustering ratio data in  {\it  scenario 1} are shown in   FIG.~\ref{fig_chi2-all_z}.
The most immediate conclusion drawn  is that the reference $\Lambda$CDM model
(the limiting case in which $|f_{R_0}|$ goes to zero) is an excellent fit to the data. The null hypothesis that the reference $\Lambda$CDM does not provide
a satisfactory description of clustering data is ruled out with a significance level of $25\%$ (for $n=2$) and $82\%$ (for $n=3$) computed  as the probability of having a $\chi^2$ statistic more extreme than
$2.77$ and $0.39$ respectively.
This result is at odds with results based on the analysis of the growth rate data which seems to favor models predicting slightly less growth of structures
than the {\it reference} $\Lambda$CDM model  \cite{sam,sbm}.

\begin{figure}
\begin{center}
\includegraphics[width=85mm,angle=0]{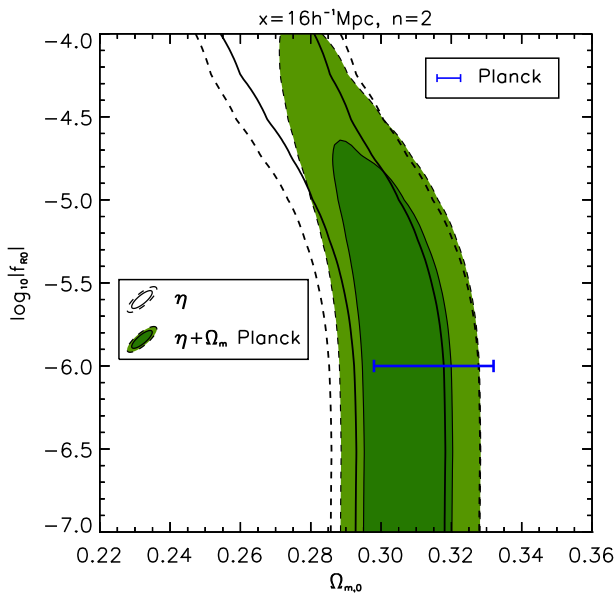}
\end{center}
\caption{Two-dimensional likelihood contours on $\Omega_{m0}$ and $f_{R_0}$ from the least square analysis of the clustering ratio $\eta(n,x)$ of the  SDSS samples $s1$ and $s2$.
The clustering ratio is estimated for  $x=16 h^{-1}$Mpc and $n=2$ in three different redshift bins ($z=0.29, 0.42$, and $0.60$).
Contours corresponds to $68$ and $95$ per cent $c.l.$ for a multivariate Gaussian distribution with $2$
degrees of freedom.  Black contours show the results obtained by fixing the baryon density $\Omega_b  h^2 = 0.0221$, the Hubble constant $ H_0 = 67.4$ km s$^{-1}$ Mpc$^{-1}$ and
the scalar spectral index $n_s = 0.96$ but letting $\Omega_{m0}$ as a free parameter. Green shaded areas show the results after implementing the Planck  Gaussian prior  $\Omega_{m0}=0.315 \pm 0.017$.
We consider  $F(R)$ models with $N=1$.
}
\label{fig_2Dlik}
\end{figure}

\begin{figure}
\begin{center}
\includegraphics[width=85mm,angle=0]{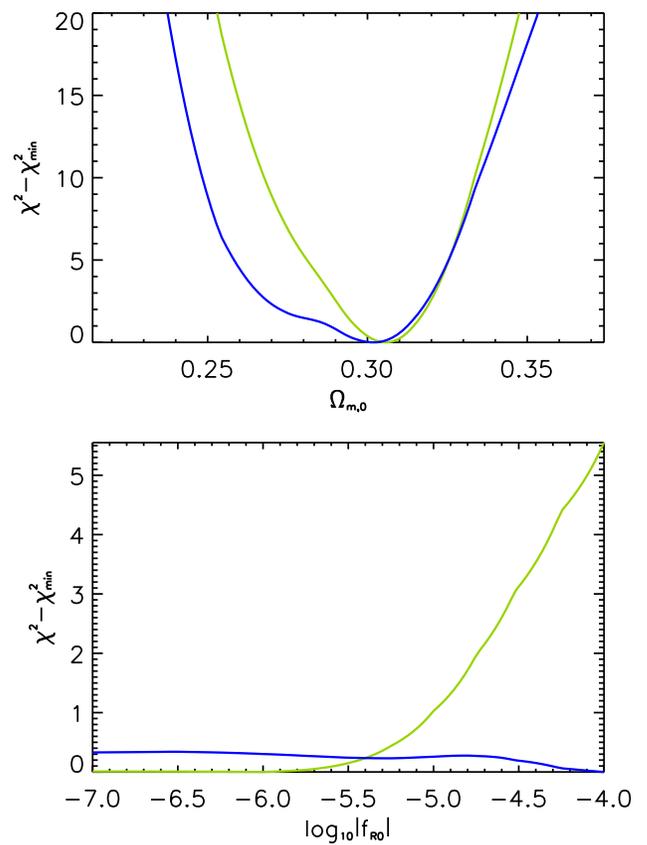}
\end{center}
\caption{Marginalised 1D likelihood  constraints on $\Omega_{m0}$ (upper panel) and $f_{R_0}$ (lower panel). The green/blue curves show the results obtained with/without the Gaussian Planck prior on $\Omega_{m0}$.}
\label{fig_1Dmarg}
\end{figure}

FIG.~\ref{fig_chi2-all_z} also shows that the smaller the correlation index $n$ the more discriminatory the $\eta$ statistic in rejecting  $F(R)$ scenarios is, essentially  because of the smaller error bars
(see FIG.~\ref{fig_etaxi_fR_x16}). Interestingly, while the reference analysis ($n=2$) provides stronger constraints,
$|f_{R_0}| \lesssim 3.2 \times 10^{-6}(/ 5.6 \times 10^{-6})$ to the $68\%$ precision level in $F(R)$ models with the exponent $N=1(/2)$
(and $|f_{R_0}| \lesssim 9.9 \times 10^{-6}(/ 1.9 \times 10^{-5})$ at the $95\%$ level), the control analysis ($n=3$), being run on different correlation scales,  allows us to check the unbiasedness
of our conclusions.
We also remark that $F(R)$ models with higher exponent N are progressively less constrained since, when compared to the N=1 models,  they display a faster convergence to the
$\Lambda$CDM model at high redshift.

Joint 2D likelihood contours on $\Omega_{m0}$ and $f_{R_0}$ obtained in {\it  scenario 2}  are displayed in FIG.~\ref{fig_2Dlik}.
The degeneracy between $\Omega_{m0}$ and $f_{R_0} $ is essentially due to the fact that the shape of the power spectrum is regulated in a similar way by these two quantities
(power is enhanced on small scales as the matter density increases or gravity becomes stronger on mildly non-linear scales).
Incidentally,  we note that this degeneracy might be somewhat alleviated if clustering ratio measurements were available
at redshifts higher  than those analysed here.  Indeed the  power spectrum becomes progressively insensitive to modified-gravity effects  at earlier epochs.

Despite this degeneracy, it is interesting to note that even allowing for modification of gravity, there is
a neat upper bound to the value of the matter density parameter that is compatible with clustering ratio measurements, specifically
$\Omega_{m0}<0.328$ at $95\%$ confidence level. This is most clearly seen in FIG. \ref{fig_1Dmarg} where  the marginalised 1D likelihood of the matter density parameter is shown. The Planck prior on the matter density  parameter does not ameliorate the already strong constraints set by the $\eta$ test for  $\Omega_{m0}>\Omega_{m0}^{\rm best\, fit}$.  Indeed the  situation is the opposite, that is, the $\eta$  constraint on $\Omega_{m0}$  improves by nearly a factor of two the precision on the matter density parameter obtained by Planck (nearly  6\% precision at 95\% c.l.).
As a consequence, all $\Lambda$CDM and $F(R)$ models
analysed in this paper with a parameter $\Omega_{m0} > 0.328$ are rejected by the
$\eta$ test alone. In other words, the upper bound on the matter density parameter obtained
within a $\Lambda$CDM model from these observations cannot be relaxed by  invoking
any  modification of gravity of the form given in Eq.~\eqref{fR-def}.

Notwithstanding,  by imposing a lower bound to the possible variation of  $\Omega_{m0}$, that is $\Omega_{m0} > 0.298$ at $68\% c.l.$
the Planck prior allows us to exclude $F(R)$ models with $| f_{R_0} |>10^{-4}$ (see FIG. ~\ref{fig_2Dlik}).  The resulting 1D constraints on $f_{R_0}$ obtained by marginalising over the matter density parameter are shown in the lower panel of  FIG. \ref{fig_1Dmarg}.
This gives $|f_{R_0}|<4.6\times 10^{-5}$  at $95\% \, c.l.$.

These results should be compared to the bounds obtained from other observables. The joint analysis of several large-scale tracers (baryon acoustic oscillations (BAO), power spectrum, lensing,  galaxy flows) combined with WMAP data gives
$B_0 < 1.1 \times 10^{-3}$ at $95\%$ $c.l.$  \cite{lomb}, where $B_0$ is defined as
$B_0= f_{RR}/(1+f_R) R' H/H'$.
This corresponds to $|f_{R_0}| < 8.4 \times 10^{-4}$, for $N=1$
[the parameterization (\ref{fR-def}) gives $B_0= -2 f_{R_0} (N+1)/(1+3\Omega_{\Lambda})$].
On cosmological scales, the best bound is $B_0 < 8.5 \times 10^{-5}$, from
 the combined likelihood of the temperature power-spectrum of Planck, of  the galaxy   power spectrum
from the wiggleZ data on scales larger than $30 h^{-1}{\rm Mpc}$, and, at lower redshift,  of the baryon acoustic oscillation (BAO) measurements from the 6dF Galaxy Survey, SDSS DR7 and   BOSS DR9  \cite{Dossett:2014oia}.
This corresponds to $|f_{R_0}| < 6.5 \times 10^{-5}$, for $N=1$.
The clustering ratio of SDSS DR10 data, being able to delve into the quasi-linear part of the power spectrum where deviations from GR are larger than on linear scales, allows one to get comparable
constraints using data from a single sample and a prior on  $\Omega_{m0}$ (from Planck).  Stronger constraints are expected when the $\eta$ statistics is combined with other gravity probes \citep{belprep}.

On smaller scales of a few kpc's, strong gravitational lensing effects of galaxies place a bound $|f_{R_0}| \lesssim 2.5\times 10^{-6}$  \cite{Smith:2009fn}, which is stronger than the one from the linear power spectrum and
of the same order as the one obtained using the clustering ratio. The absence of disruption of the dynamics of satellite galaxies of the Milky Way implies that the latter must be screened, implying a loose bound of
$| f_{R_0}| \lesssim 7\times 10^{-7}$ \cite{Lombriser:2014dua}.

Effects on distance indicators in dwarf galaxies \cite{Jain:2012tn}, and the comparison between the gas and stellar dynamics in these galaxies \cite{Vikram:2013uba} imply that $|f_{R_0}|\lesssim 5\times 10^{-7}$. Finally, the most severe constraint in the Solar System comes from the test of the strong equivalence principle by the Lunar Ranging experiment \cite{Williams:2012nc} at the $10^{-13}$ level, which results in a competitive bound $|f_{R_0}|\lesssim 10^{-6}$ for $N=1$ and irrelevant ones for greater values of $N$.

All in all, we find that the clustering ratio provides a method to test the properties of modified gravity almost
as sharp as Solar System experiments such as Lunar Ranging or strong lensing observations, and better than current observations of the growth of cosmological structures on linear scales. Only dwarf galaxies where the chameleon effects are enhanced between screened and unscreened objects are more discriminatory.

\begin{figure}
\begin{center}
\includegraphics[width=85mm,angle=0]{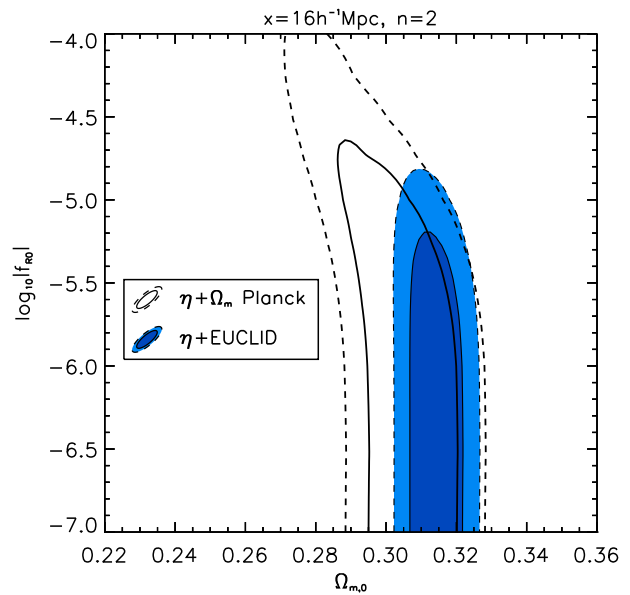}
\end{center}
\caption{Two-dimensional likelihood contours on $\Omega_{m0}$ and $f_{R_0}$ from the least square analysis of the clustering ratio $\eta(n,x)$ of the  SDSS samples $s1$ and $s2$.
The clustering ratio is estimated for  $x=16 h^{-1}$Mpc and $n=2$ in three different redshift bins ($z=0.29, 0.42$, and $0.60$).
Contours corresponds to $68$ and $95$ per cent $c.l.$ for a multivariate Gaussian distribution with $2$
degrees of freedom.  Black contours show the results obtained by fixing the baryon density $\Omega_b  h^2 = 0.0221$, the Hubble constant $ H_0 = 67.4$ km s$^{-1}$ Mpc$^{-1}$ and the scalar spectral index $n_s = 0.96$ but letting $\Omega_{m0}$ as a free parameter. Blue shaded areas show the results after combining with the expected constraints from Euclid. We consider  $F(R)$ models with $N=1$. }
\label{likeuclid}
\end{figure}

Looking into the future, we have used the  Horizon  mock surveys \cite{hor} to extrapolate some
forecasts for the errors on $\eta$  achievable by Euclid, a future redshift survey with characteristics similar to
SDSS, but covering larger and deeper space volumes.
Computations in {\it scenario 1}, {\it i.e.} by assuming the precise knowledge of the background cosmology, show that such a mission will be able to push the
statistical error on measurements of $\eta$  at $z \sim 1(/1.5)$ below $0.9\% (/1.1\%)$  (we consider $n=2$ and $x=16 h^{-1}$Mpc).
We find that, although the clustering ratio in $F(R)$ scenarios significantly differs from that expected in the $\Lambda$CDM model only at late epochs,  when cosmic acceleration kicks in,
high redshift Euclid  measurements are expected to lower the $95\%$  bound on $f_{R_0}$ by roughly a factor of $4$. Therefore, even in the near future,
cosmological constraints on $F(R)$ gravity  are not expected to improve  on astrophysical bounds. This result is specific to models with the chameleon property. The analysis of alternative screening mechanisms
like K-mouflage \cite{Brax:2014yla},   where large objects such as galaxy clusters are not screened,  will certainly make  Euclid-like data more discriminatory. A similar improvement  is seen when
allowing for uncertainties in  the knowledge of the matter density parameter ({\it scenario 2}). Indeed  FIG. \ref{likeuclid} shows that the combination of high (EUCLID) and low (SDSS) redshift
estimates of the galaxy clustering ratio will allow to break the degeneracy between $f_{R_0}$ and $\Omega_{m0}$. As a result, the $95$\% upper bound on $f_{R_0}$ is expected to decrease down to $\sim  10^{-5}$, nearly a factor $5$  improvement on current constraints.

\section{Conclusion}
\label{sec:Conclusion}

We have extended the clustering ratio method to the study of modified gravity models in their cosmological regime and in the quasi-linear regime of structure formation. We have shown that the accuracy of the comparison between the theoretical calculation of the
clustering ratio and data reaches 2\%. Using the $F(R)$ models in the large curvature regime as a template for modified gravity, we find that this is enough to
obtain competitive bounds on parameters such as $f_{R_0}$ when the matter fraction is fixed or allowed to vary within a prior given by Planck. In the first case, we find that the bound on $|f_{R_0}| \lesssim 3 \times 10^{-6}$ (at the $68\%$ confidence level) is of the same order as
the one from the Solar System. This also gives a $10^{-5}$ bound at the
$95\%$ confidence level.
 In the second case, the $95\%$ {\it c.l.}  bound becomes
$|f_{R_0}|\lesssim 4.6\times 10^{-5}$. This is slightly better than the cosmological limit obtained in the linear regime of perturbation theory.

More precisely, having assumed that Planck measurements provide an  accurate mapping of redshifts into distances, i.e a precise description
of the smooth expansion rate history of the Universe, we have shown  that the reference $\Lambda$CDM model
describes  the linear clustering properties of SDSS galaxies in the redshift range $0.15<z<0.67$, that is Einstein's General Relativity satisfactorily describes also the perturbed
dynamics  of the late Universe. In particular, by fixing the relevant cosmological parameters to the Planck central value, $F(R)$  models having the same expansion rate as
the reference $\Lambda$CDM model are  excluded at $95\%$ by the $\eta$-test of gravity
if $|f_{R_0}| >10^{-5}$ (if $\Omega_{m0}$ is fixed) and  $|f_{R_0}| >4.6\times  10^{-5}$ (if we have a Planck Gaussian prior on $\Omega_{m0}$).
Based on this encouraging result, an extensive  likelihood analysis  is  being conducted with the aim of using the $\eta$ statistic to assess the viability  of a more general class of modified gravity models, such as dilaton and symmetron models
\cite{BV1,BV2},
which can exhibit greater deviations from $\Lambda$CDM, or $K$-mouflage models
\cite{BV-K1,Brax:2014yla},
where both the background and the perturbations deviate from $\Lambda$CDM.

\begin{acknowledgments}

We acknowledge useful discussions with  L. Perenon and  F. Piazza.
CM is grateful for support from specific project  funding of the Institut Universitaire de France and of the Labex OCEVU.
JB acknowledges support of the European Research Council through the Darklight ERC Advanced Research Grant (\#291521).
This work is supported in part by the French Agence Nationale de la Recherche under Grant ANR-12-BS05-0002. P.B.
acknowledges partial support from the European Union FP7 ITN
INVISIBLES (Marie Curie Actions, PITN- GA-2011- 289442) and from the Agence Nationale de la Recherche under contract ANR 2010
BLANC 0413 01.

\end{acknowledgments}

\end{document}